# Deterministic Storage of Quantum Information in the Genetic Code

Roberto Rivelino*

Instituto de Física, Universidade Federal da Bahia, 40210-340 Salvador, Bahia, Brazil

**ABSTRACT:** DNA has been proposed as a chemical platform for computing and data storage (Yang, S., *et al.*, *Nat. Rev. Chem.* **2024**, *8*, 179), paving the way for building DNA-based computers. Recently, DNA has been hypothesized as an ideal quantum computer with the base pairs working as Josephson junctions (Riera Aroche, R., *et al*. *Sci. Rep.* **2024**, *14*, 11636). There are still major challenges to be overcome in these directions, but they do not prevent deviceful perspectives of the main problem. The present paper explores DNA base pairs as elementary units for a scalable nuclear magnetic resonance quantum computer (NMRQC). First, it presents an overview of the proton transfer (PT) mechanism during the prototropic tautomerism in the base pairs, scoring the current stage. Second, as a proof-of-principle, the paper examines these molecular structures as quantum processing units (QPUs) of a biochemical quantum device. For the model proposed here, it is theoretically demonstrated that the nuclear spins involved in the PT of base pairs can be deterministically prepared in a superposition of triplet states. Under appropriate conditions, the proton dynamics provides the minimal two-qubit entanglement required for quantum computing. The dynamics between the canonical and tautomeric quantum states (CQS and TQS, respectively) is determined from a thermally dependent Watson-Crick quantum superposition (WCQS); i.e., $|\text{WCQS}\rangle = a(T)|\text{CQS}\rangle + b(T)|\text{TQS}\rangle$ with $|a(T)|^2 + |b(T)|^2 = 1$. If the DNA structure is sufficiently protected to avoid environment-induced decoherence of the confined-proton quantum states, quantum information can be successfully encoded in several base pairs along the coiled double strand. As a potential applicability, a crystalline DNA device could be employed for quantum computing and cryptography controlled by a sequence of Ramsey pulses. Finally, this study critically evaluates these possibilities toward a proof-of-concept of a DNA-based quantum computer.

# 1. INTRODUCTION

The genetic information responsible for the structure, function, and development of all living organisms and viruses is encoded in polymeric macromolecules, namely deoxyribonucleic acid (DNA) and ribonucleic acid (RNA). These molecules are usually located in the cell nucleus compacted into chromosomes,[1] except for viruses that exhibit more 'simplified' versions of a genome[2] compared to that of eukaryotic cells.[3] The building blocks of DNA or RNA are nucleotides, made up of a sugar molecule (deoxyribose in DNA, and ribose in RNA), a phosphate group, and the nucleobases: adenine (A), guanine (G), thymine (T), and cytosine (C) in DNA, with uracil (U) substituting T in RNA (see Figure 1). Thus, strings of nucleotides bonded in specific sequences (genes) are employed for storing and transmitting heredity from proper protein syntheses.[4] These findings, combined with modern sequencing techniques,[5,6] summarize the basic of molecular biology for understanding the genetic code running on Earth *ca*. 3.5 billion years.[7,8]

DNA exhibits the most emblematic mechanism of encoding genetic instructions for all forms of life, such that a complete nucleotide sequence represents the entire genetic endowment of the organisms. The common structure of DNA consists of two strands coiled around each other (Figure 1) forming a double-helix macromolecule, albeit other DNA sequences may adopt multistranded structures (e.g., triplex and quadruplex) under physiologic conditions.[9,10] From the chemical viewpoint, the DNA framework is essentially maintained held together by hydrogen bonds (H-bonds) between nucleotides, and π-base-stacking interactions involving the nucleobases A, G, T, and C.[11,12] The Watson-Crick combinations between two DNA strands are A–T and G–C, which gives rise to a kind of canonical complementarity.[13] The reason for this comes from the possibility of forming specific H-bonds between the base pairs.[14]

Because of its highly compacted stable structure, and capacity of encoding biological information, DNA has been proposed as a promising data storage material.[15-17] From the viewpoint of the classical information theory, a native DNA molecule can encode, for example, two binary bits per nucleotide (A, T, C, and G).[18] This means that theoretically each nucleotide can store 2-bit information that has twice the capacity of conventional storage media. Thus, it is possible to store an enormous amount of information in a relatively small space in DNA molecules.[17] In this sense, synthetic DNA has emerged as a superlative medium to efficiently encode computer data[19] with accurate capacity of the readout processing.[20] However, the development of error-correction coding in DNA-based data storage remains a challenge.[21]

In addition to the usual composition of DNA, the nucleobases may exist in several isomeric forms,[22] consequence of their keto-enol or amino-imino prototropic tautomeric equilibria. Upon the formation of A–T or G–C base pairs, the preferential tautomeric forms also depend on the external environment of a DNA molecule, such as the pH of the medium[23-25] or chemical modifications.[26,27] In native DNA, the nitrogen atoms attached to A, G, and C are in the amino form (–$NH_2$), whereas the oxygen atoms attached to G, C, and T are in the keto form (C=O).[28] Considering the tautomeric equilibrium of the base pairs within the Watson-Crick model of DNA,[29] one obtains the imine tautomers of A and C, denoted by A* and C*, respectively; and the enol tautomers of G and T, denoted by G* and T*, respectively. In this sense, taking into account the formation of new H-bonds after DNA replication, other tautomeric nucleobase combinations, e.g., A*–C, A–C*, G*–T, and G– T*, are associated to spontaneous point mutations.[28-30]

This perspective paper gives first an overview of the PT mechanisms in the prototropic tautomerism of DNA base pairs, looking for a quantum protonic dynamics and discussing the implications of quantum and thermal effects on the process. As a proof-of-principle, it examines theoretically these molecular structures as elementary units of quantum information. From the proposed quantum model, it is demonstrated that the nuclear spins involved in the PT of the base pairs are entangled and form a deterministic superposition of triplet states. Thus, the proton dynamics provides the minimal two-qubit entanglement required for quantum computing. The canonical quantum state (CQS) and the tautomeric quantum state (TQS) of a base pair are investigated from a thermally dependent Watson-Crick quantum superposition (WCQS) between the CQS and TQS. The viability of the model is analyzed within the zero-field splitting (ZFS) interaction. Furthermore, schemes are proposed to simulate the proton dynamics in the base pairs via quantum computing. Potentially, DNA can be employed to create controlled quantum gates using appropriate sequence of Ramsey pulses.

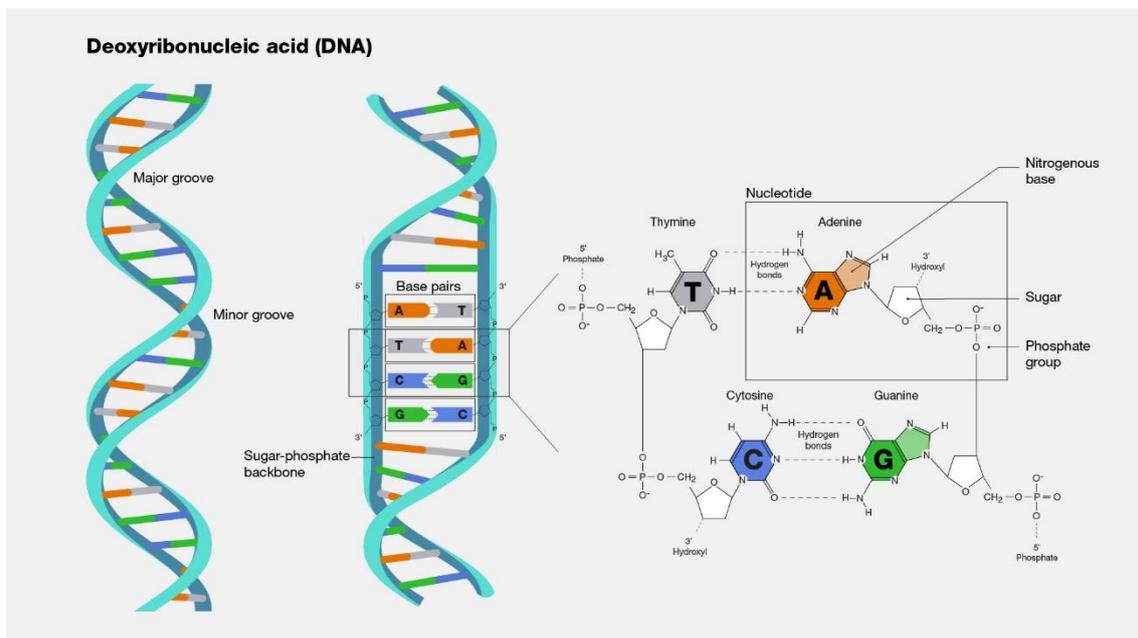

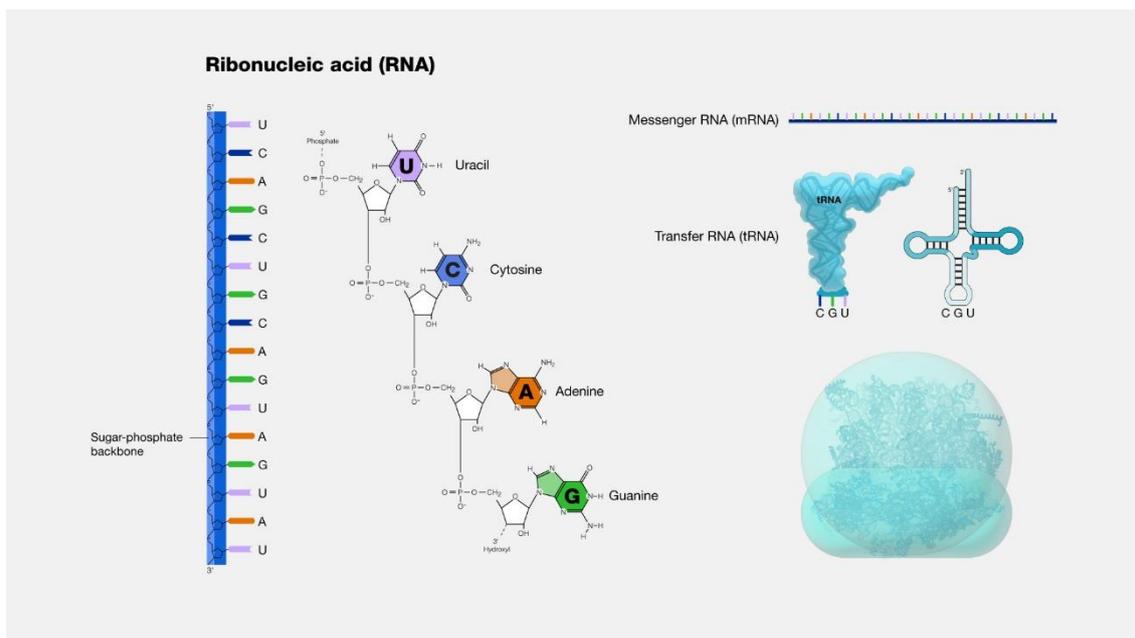

**Figure 1**. DNA (top) and RNA (bottom) structures. Courtesy: National Human Genome Research Institute. https://www.genome.gov

## 2. AN OVERVIEW ON THE PROTON TRANSFER MECHANISM IN DNA

### 2.1. Mapping the Prototropic Tautomerism in DNA Base Pairs

Starting with the canonical forms of the base pairs within the Watson-Crick picture (Figure 2), the products A*–T* and G*–C* result after a distinguishable transfer of $H_3$ in T to A (with back transfer of $H_6$ in A to T) and $H_1$ in G to C (with back transfer of $H_4$ in C to G). Notwithstanding, $H_2$ in G prefers to remain in a localized bound state during the tautomeric equilibrium.[28,30] (See the atomic labels in Figure 2). In this classical picture, the discussion about whether the double PT (DPT) mechanism is synchronous (SDPT) or asynchronous (ADPT) is an issue of intense debate.[25,31-34] Indeed, upon the prototropic tautomerization, it is possible that the complementarity between the base pairs is altered. Hence, the PT mechanism may influence the genetic information, leading to a natural error (due to mismatching) in the first cell duplication.[29]

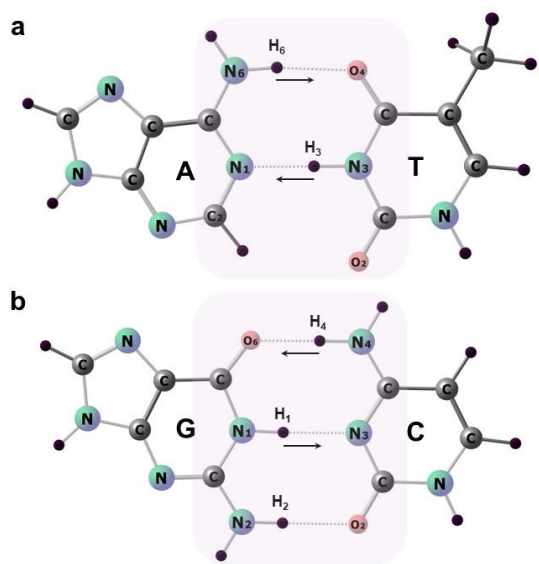

**Figure 2**. Classical picture of the PT mechanism in DNA base pairs. Canonical structures of (a) A–T and (b) of G–C. Shaded boxes indicate the intermolecular H-bonding region and black arrows indicate the typical prototropic tautomerism in these base pairs.

The prototropic tautomerism of A–T and G–C has been an issue of broad and permanent interest because of its biological relevance to understand mutation.[35-46] Most of these studies have treated tautomerism classically, with some attention to the quantum description of the proton motion.[38,47,48] Commonly, the energy landscape of the canonical and tautomeric forms of DNA base pairs has been modeled using electronic structure calculations combined with the transition state theory (TST).[49,50] Thus, from the resulting potential energy profile, the PT can be investigated trough hopping or tunneling of protons leading to A*–T* and G*–C* (see Figure 2). Slocombe and co-workers have carefully examined this problem in recent papers.[46,48,51,52]

By employing a pseudo-one-dimensional pathway for the PT in the canonical G–C and A–T bases pairs, Slocombe *et al.*[51] have analyzed quantum and classical effects for the point mutations of DNA. Assuming that the two protons in a base pair undergo asynchronous transfer, these authors[48] have proposed that the quantum tunneling contribution is determining for the process. Quantum corrections significantly shorten the classical rate estimates, indicating that the protons are sufficiently delocalized along the H-bond and populate both states of the base pairs. During DNA replication, the tautomeric states A*–T* and G*–C* might play a relevant role for spontaneous mutations by copying the wrong pairs A*–C, A–C*, G*–T, or G– T* from native DNA.[52,53] Thus, if the PT is dominated by a quantum mechanism, the basic unit of a genetic message will ultimately be also subject to quantum effects.

**2.2. Thermal and other External Effects on the PT Mechanism**

The structure of a DNA molecule is sensitive to thermal[54] as well as other external effects, such as solvents[24,25,55-57] and electric fields.[58,59] Furthermore, its helical double-strand structure can be destabilized by increasing temperature, leading to the full separation of the two strands.[60] Consequently, from low to moderately high temperatures, the PT process in DNA base pairs is also thermally dependent.[61] The motion of the hydrogen atoms is faster than that of heavier atoms to which they are covalently bonded; then, when an H-bond is formed their degree of vibrational motion becomes restricted, since two bonds instead of one restrain the H atom.[62] Hence, the vibrational stretching of the covalently bonded H undergoes the largest frequency change and provides insight into the potential energy surfaces of H-bonds (see Figure 3).

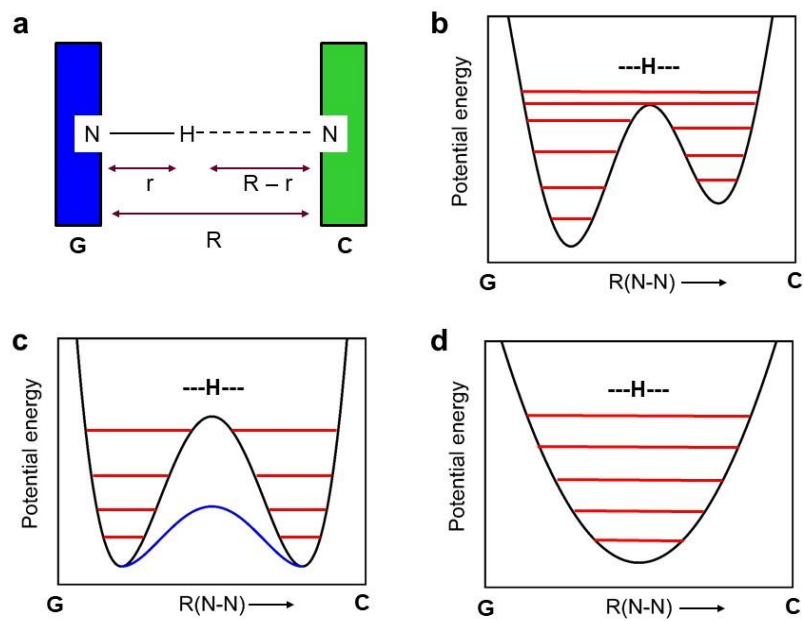

**Figure 3**. (a) One-dimensional model for the potential energy curve (PEC) of a proton (H$^+$) in an H-bonded base pair during the tautomeric equilibrium. (b) When a moderate H-bond is formed, the PEC becomes broader and a second minimum appears, leading to an asymmetric double minimum potential. (c) For stronger H-bonds, the two minima become more symmetric, and the potential barrier reduces (blue curve) forming a low-barrier H-bonding. (d) In some conditions, the proton transfer accompanies the formation of ionic H-bonds, leading to a PEC with a single minimum.

Considering as a simple example the N–H···N bond present in the G–C pair (Figure 3a), the PEC of a covalently bonded H-atom develops a second minimum, and the vibrational levels of the N–H bond stretching become closer (red lines in Figure 3b).[48] Depending on modifications in each nitrogenous nucleobases, the PEC barrier can be reduced,[41] the two minima can become more symmetric (Figure 3c), and in extreme cases the barrier can even disappear (Figure 3d). However, the simplest PT process in a canonical base pair does occur via a certain mechanism involving at least a DPT profile.[47] Winokan *et al.*[63] have conducted an interesting discussion of the PT mechanism involving the formation of zwitterionic states.

The main issue here is to know whether the proton distribution in DNA base pairs under different temperatures arises from a classical statistical mixture of tautomers or from a quantum mechanical superposition between the CQS and TQS.[64] Additionally, it is important to make sure that the system is in a coherent superposition of macroscopic tunneling states, in order to demonstrate some macroscopic quantum correlation.[65] Indeed, thermal effects in the prototropic tautomrism can be assessed using different theoretical models[48,66-70] involving proton dynamics. However, despite quantum-mechanical corrections being applied to the motion of protons in the tautomerism, a quantum mechanism for the proton dynamics remains incomplete.

## 3. THE PROPOSED MODEL AND ITS THEORETICAL APPROACH

### 3.1. Quantum States of Protons in DNA Base Pairs

The quantum effects on the PT mechanism of H-bonded systems is unusually large when the effective mass of the 'conduction' $H^+$ is smaller than the mass of a bare proton ($\frac{m_{H^+}}{m_p} < 1$).[71] For this reason, the dynamics of protons involved in H-bonding is greatly influenced by their quantal behavior even at moderate temperatures,[72,73] in nanoconfined environment,[74] or under high pressures.[75] Interestingly, the momentum distribution of protons in condensed water phases is significantly different from the classical equilibrium distribution.[76,77] In this sense, one expects that the proton correlation is also important inside a DNA molecule and, consequently, that the quantum state of protons may be entangled in an H-bonded base pair.

For a system containing multiple H-bonds in a spatially confined region, such as DNA as well as other systems,[74,78] a many-proton state function can be properly proposed.[79-81] However, the partitioning scheme for considering the quantum protons is not universal. A consensual fact is that the protons involved in the tautomeric process are delocalized particles, and quantum effects can change the relative stability of the tautomers.[48] Hence, the PT dynamics can be adiabatically separated from the heavy atoms of the nucleotides, resulting in a superposition of oscillations of protons between the two configurations takes place via quantum beats.[64] A reasonable starting point to encompass the tautomeric process is considering a fully-quantum-mechanical picture for the confined protons in each base pairs of DNA.

Taking into account the Hilbert subspace of the protons involved in the tautomeric equilibrium, the two-particle quantum states of A–T and G–C can be written in the coordinate representation[64] as the following quantum superposition:

$$\Theta_{XY}^{WC}(x_1, x_2) = a(T)\Phi_{XY}(x_1, x_2) + b(T)\Phi_{X^*Y^*}(x_1, x_2) \qquad (1)$$

Here, it is necessary to give a physical meaning and define the labels for eq 1. The superscript WC indicates that $|\Theta_{XY}^{WC}\rangle$ is a Watson-Crick two-proton quantum state (WCQS), while the subscript XY (or X*Y*) indicates a specific canonical X–Y base pair (or its tautomeric X*–Y* counterpart). In this adiabatic partitioning, $|\Phi_{XY}\rangle$ is a pure canonical two-proton quantum state (CQS) while $|\Phi_{X^*Y^*}\rangle$ is its tautomeric counterpart (TQS). As already observed for related systems,[64] eq 1 is defined as a quantum superposition in thermal equilibrium of the CQS and TQS, satisfying the probability constraint $|a(T)|^2 + |b(T)|^2 = 1$, with $|a(T)|^2 \gg |b(T)|^2$. For example, at $T = 300$ K, the calculated tautomer occupation probability is of $1.73 \times 10^{-4}$ (see ref 48). Finally, the coordinates $x_i \equiv \{r_i, s_i\}$, with $r_i = (x_i, y_i, z_i)$ being a proton spatial position and $s_i = \pm 1/2$ being its spin component ($i = 1,2$). These coordinates (position and spin) are depicted in Figure 4 for a G–C base pair model.

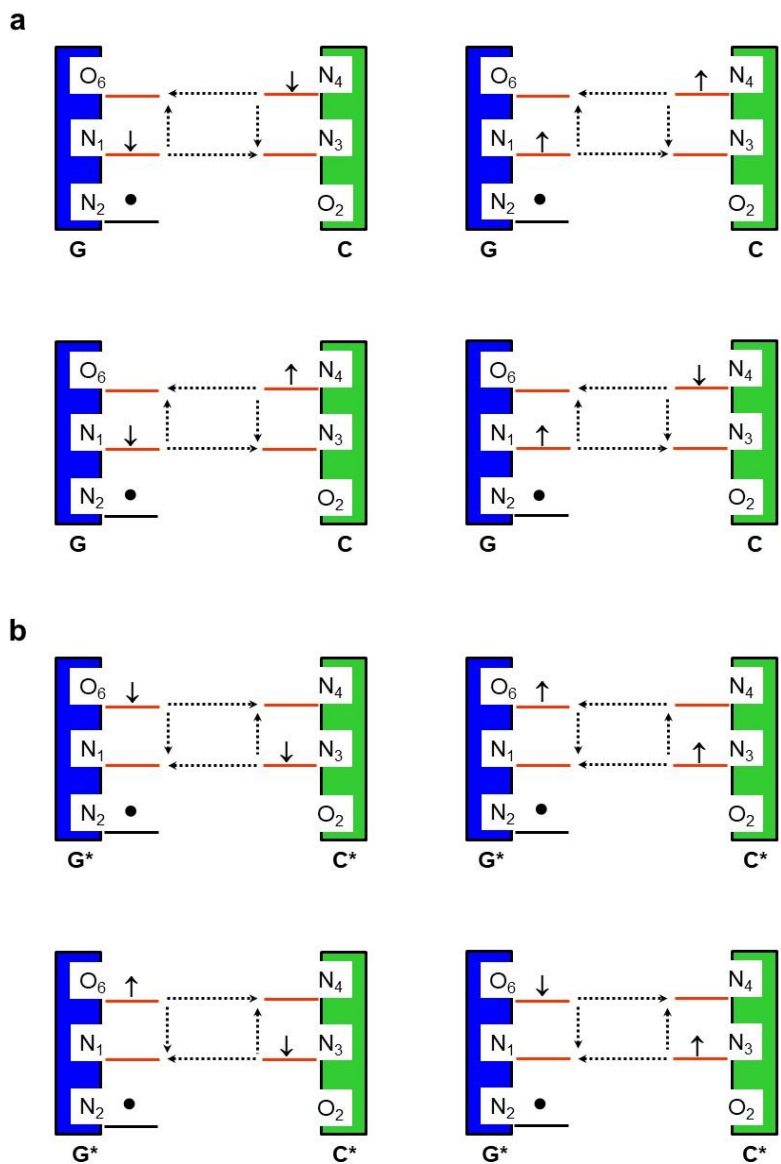

**Figure 4**. Possible spin configurations in a weakly interacting two-proton model (WI2PM) for (a) the canonical G–C base pair leading to a CQS and (b) the tautomeric G*–C* base pair leading to a TQS. Spins up (↑) and down (↓) of the protons are represented as one-particle independent spin-states. The proton in $N_2$ (•) is not directly involved in the usual tautomerization process.

**3.1.1. Adiabatic partitioning of the system**. One assumes that at low temperatures, and specifically at $T = 0$ K, a system described by eq 1 is in a quantum superposition and the protons are spatially delocalized. For a DNA at higher temperatures, the WCQS is, in principle, permitted, although the environment-induced decoherence is supposed to lead rapidly the system to a statistical mixture of the two configurations, thus avoiding quantum interference.[82] However, there is experimental evidence for macroscopic coherent tunneling of protons in H-bonded crystals[64] as well as in DNA base pairs.[48] Therefore, in the absence of boundary decoherence, there might be a nonlocal proton dynamics with quantum entanglement. In practice, the double strand structure of DNA works as a confining wall for the protons involved in the internal H-bonds. With these general arguments, the proton motion is adiabatically separated from the heaviest atoms, resulting in a confined system of weakly interacting particles, as illustrated in Figure 4.

Using this approach, the Hamiltonian of an X–Y base pair can be written in the form

$$\mathcal{H} = \mathcal{H}_Q^{H^+} + \mathcal{H}_{BO}^{XY} + \mathcal{H}_{coup}^{XY-H^+} \qquad (2)$$

where, $\mathcal{H}_Q^{H^+}$ is a quantum Hamiltonian for the dressed or effective protons (H$^+$), containing the kinetic energy and Coulomb repulsion, $\mathcal{H}_{BO}^{XY}$ is the conventional molecular Hamiltonian (e.g., in the Born-Oppenheimer approximation) for describing the electronic structure of the X–Y pair, without the H-bonded protons, and $\mathcal{H}_{coup}^{XY-H^+}$ is the coupling between both subsystems. For the present construction, it is not necessary to model this latter term now. In order to proceed any further with this problem, or with a more sophisticated approach, it is important to evaluate matrix elements of eq 2 between Slater determinants, since the effective protons are identical fermionic particles that carry their spin ½.

**3.1.2. The two-proton interaction model.** As a first approximation, it is interesting to consider a weakly interacting two-proton model (WI2PM) depicted in Figure 4 as an independent particle model (IPM). Alternatively, the WI2PM could be represented by a charge resonance model.[83] Briefly, it is possible to write a CQS (or TQS) in terms of a single determinant; then, for the G–C base pair, the approximate CQS is of the form

$$\Phi_{GC}(x_1, x_2) = \frac{1}{\sqrt{2}} \begin{vmatrix} \chi_i^{GC}(x_1) & \chi_j^{GC}(x_1) \\ \chi_i^{GC}(x_2) & \chi_j^{GC}(x_2) \end{vmatrix} \qquad (3)$$

In eq 3, $\chi_i^{GC}$ and $\chi_j^{GC}$ are delocalized single-proton spin-space functions belonging to the G–C pair. It is also reasonable to assume that an arbitrary spin-space function is of the form $\chi = \psi(r)\xi_m(s)$, with $\psi(r)$ being a single-proton space function and $\xi_m(s)$ representing a single-proton spin state. To some extent, the one-proton space function can be simulated by forming linear combinations, i.e., $\psi_+(r) \approx \varphi^G(r) + \varphi^C(r)$ or $\psi_-(r) \approx \varphi^G(r) - \varphi^C(r)$. In this approximation, $\varphi^G(r)$ and $\varphi^C(r)$ are spatially localized single-proton functions in G and C nucleotides, respectively, which fit well into the Watson-Crick model.

To gain more insight into the proposed two-proton quantum state of G–C, it is mandatory to analyze the role of the proton exchange in the stabilization of the base pair starting with the WI2PM. Considering the total spin state for the protons involved in the tautomeric equilibrium a posteriori, the total space function can be written now in the form

$$\Psi_{GC}^{(\pm)}(r_1, r_2) = \frac{1}{\sqrt{2}}[\psi_i^{GC}(r_1)\psi_j^{GC}(r_2) \pm \psi_j^{GC}(r_1)\psi_i^{GC}(r_2)] \qquad (4)$$

For an IPM, the two products involving the spatial functions in rhs of eq 4 are energetically degenerate. However, the proton-proton repulsion present in $\mathcal{H}_Q^{H^+}$ is sufficient to remove this degeneracy, resulting in an energy gap between these two protonic states. In this model, the proton exchange role can be understood from the amplitude probabilities of eq 4 as one proton approaches the other (see Figure 5a); that is $\left|\Psi_{GC}^{(-)}(r_1 = r_2)\right|^2 = 0$ whereas $\left|\Psi_{GC}^{(+)}(r_1 = r_2)\right|^2$ is maximal. In other words, there is a null probability of finding two protons in the same infinitesimal space region inside a base pair before the tautomerism starting to occur, and then the ground state of G–C might be described by $\Psi_{GC}^{(-)}$ instead of $\Psi_{GC}^{(+)}$ (see Figure 5).

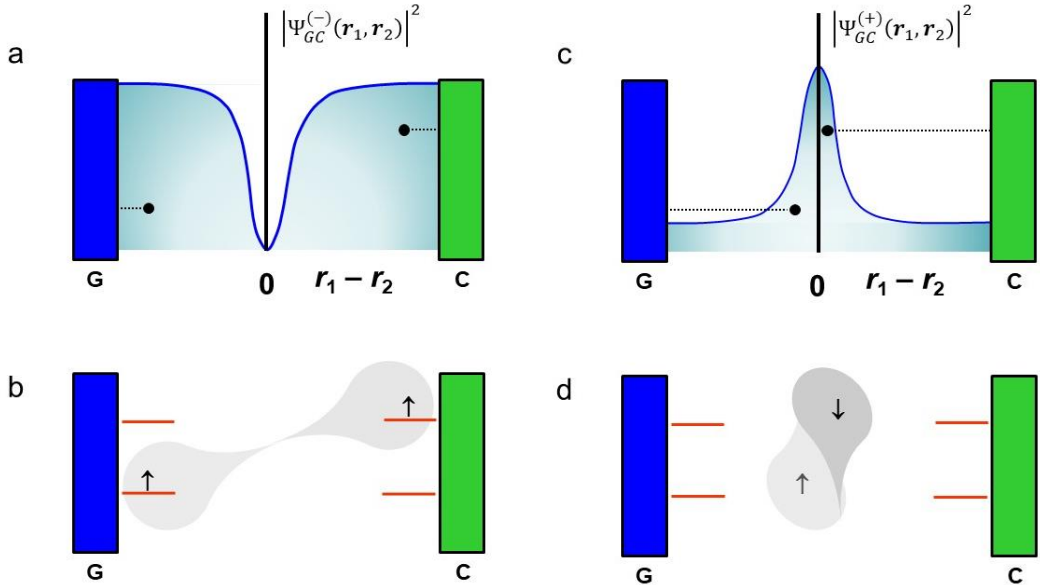

**Figure 5**. Correlated spatial probability densities of two protons in a DNA base pair and their spin state counterparts. (a) The dip in the probability density $\left|\Psi_{GC}^{(-)}\right|^2$ when $r_1 \approx r_2$ is a Fermi hole, which is a purely quantum mechanical phenomenon, and does not depend on the Coulomb repulsion of the protons. (b) A possible spin triplet for the CQS when the proton spins are unpaired. (c) The maximum in the probability density $\left|\Psi_{GC}^{(+)}\right|^2$ when $r_1 \approx r_2$ indicates a Fermi heap that occurs during the PT mechanism. (d) A possible spin singlet for an excited CQS corresponding to the classical transition state (TS).

The protonic excited state of the canonical G–C base pair in this two-proton model should be thus $\Psi_{GC}^{(+)}$ (Figure 5c). Next, to obtain the ground and excited CQS, it is necessary multiplying $\Psi_{GC}^{(-)}(r_1, r_2)$ and $\Psi_{GC}^{(+)}(r_1, r_2)$, respectively, by proper two-proton spin states (see Figure 5b and 5d) in order to ensure the antisymmetry principle. It is clear herein that these spin states must be symmetric or triplet state ($S = 1$) for $\Psi_{GC}^{(-)}$ and antisymmetric or singlet state ($S=0$) for $\Psi_{GC}^{(+)}$. These are designated by using the $\{|S, M_S\rangle\}$ basis in the WI2PM Hilbert subspace, as

$$\xi_{M_S}^{(S=1)}(s_1, s_2) = \begin{cases} |1,1\rangle = |\uparrow\uparrow\rangle \\ |1,0\rangle = \frac{1}{\sqrt{2}}(|\uparrow\downarrow\rangle + |\downarrow\uparrow\rangle) \\ |1,-1\rangle = |\downarrow\downarrow\rangle \end{cases} \tag{5}$$

$$\xi_{M_S}^{(S=0)}(s_1, s_2) = |0,0\rangle = \frac{1}{\sqrt{2}}(|\uparrow\downarrow\rangle - |\downarrow\uparrow\rangle) \tag{6}$$

In this model, the ground CQS becomes $\Phi_{XY}(x_1, x_2) = \Psi_{GC}^{(-)}(r_1, r_2)\xi_{M_S}^{(S=1)}(s_1, s_2)$ and the excited CQS becomes $\Phi_{XY}^{\ddagger}(x_1, x_2) = \Psi_{GC}^{(+)}(r_1, r_2)\xi_{M_S}^{(S=0)}(s_1, s_2)$, in the coordinate representation.

**3.1.3. Reactant-product coherence.** It is important to highlight that the 'quantum H-bonds' involving two protons, such as those present in DNA base pairs, dramatically differ from an electronic chemical bond, such as that in the H$_2$ molecule, which leads the electronic ground state to be a singlet and the antibonding state to be a triplet.[84] However, in the case of a base pair, the spatial antisymmetry needs to be assumed to connect the quantum dynamics between the CQS and TQS.[85] Although the protonic picture seems like only an appealing model, it is consistent with the classical description of the transition state theory (TST),[49] which assumes a special type of quasi-equilibrium between the reactant (G–C) and an activated TS complex (G$^{\ddagger}$–C$^{\ddagger}$).

Based on the reactant-product coherence assumption, when the tautomerism starts to occur, it becomes clearer that the proton excitation evolves from a long-lived triplet state (the reactant) to a short-lived singlet state (the TS complex). Equivalently, the TQS of G*–C* should be also a triplet state, up to a global phase factor, connecting to the same TS complex during the prototropic tautomerism (see Figure 6). Thus, considering the tautomeric equilibrium G–C ($S = 1$) ↔ G$^{\ddagger}$–C$^{\ddagger}$ ($S = 0$) ↔ G*–C* ($S = 1$), one singlet excited state can be rapidly converted in a superposition of triplet states ($S^{\ddagger}_1 \rightarrow T_0 + T_0^*$), through a mechanism that resembles the singlet fission,[83,86] but applied to the protons[87] instead of electrons. This is better clarified in the following.

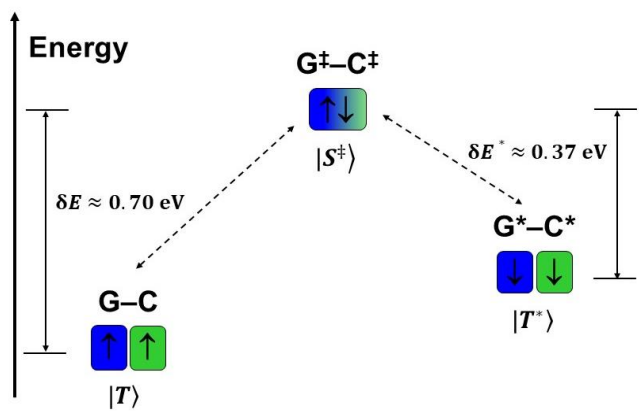

**Figure 6**. The singlet fission model for the PT mechanism in G–C. The values of $\delta E$ ($\delta E^*$) are taken from ref 48. In fact, considering these values one obtains protonic transitions in the range of typical vibrational frequencies, which are compatible with the present quantum model.

This spin-allowed process indicates that the singlet state can be safely neglected thereby decaying into two stable triplets, and satisfying the WCQS superposition described by eq 1. During the tautomeric equilibrium, the very same state could also mediate the reverse of singlet fission; i.e., a triplet–triplet annihilation (TTA), where two triplets could fuse to form the TS singlet.[83] For electronic excitations, such a process has been exploited for high-density secure data storage[88] and quantum information.[89] A simple analysis here shows that, for $\delta E$ (considered as the forward reaction barrier[48] of ~0.7 eV and displayed in Figure 6) connecting G–C to G$^{\ddagger}$–C$^{\ddagger}$, the singlet excited state decoherence time ($\tau_s \geq \hbar/\delta E$) is about ~1 fs. While electronic triplet-pair states and inter-triplet interactions appear to be elusive,[90] the situation of spin-entangled triplet-pair state is very compatible with the delocalized protons during the tautomeric equilibrium.

Moreover, this reversible tautomeric reaction could be also controlled by a specific photophysical mechanism.[68]

Within this model, it is still possible to employ a quasiclassical approach[91] to estimate the characteristic PT time over the distance $R \approx 2.86$ Å in G–C (see Figure 3a). This time can be calculated by using the formula $\tau_{PT} = \nu^{-1} e^{R/r}$.[92] Thus, taking into count that the vibrational stretching ($\nu$) of N–H bonds are in the 3150–3300 cm$^{-1}$ range,[93] and considering the localization radius ($r$) as the equilibrium distance of the N–H bond in G–C, one obtains $\tau_{PT} \approx 0.3$ ps. This is much longer than the singlet decoherence time ($\tau_s \approx 1$ fs), and is comparable to the momentum relaxation time of the proton in ultrafast intramolecular tautomerism.[94] This estimate indicates that the PT mechanism in this type of tautomeric equilibrium can really occur within deterministically entangled triplet-pair states in the context of NMR.[87] This is analyzed further in detail treating the small terms in the protonic Hamiltonian.

### 3.2. Small Terms in the Protonic Hamiltonian of a Base Pair

**3.2.1. Direct spin-spin interaction.** It is well known that the three triplet spin components are non-degenerate, even in the absence of a magnetic field. Hence, considering the zero-field splitting (ZFS) for the protons in a DNA base pair, the states are determined by magnetic interactions such as direct spin-spin (SS) and less likely spin-orbit (SO) interactions. In the first case, considering the magnetic dipole-dipole interaction between the unpaired protons, which make up the triplet states, the SS term is governed by the Hamiltonian

$$\mathcal{H}_{SS} = \gamma^2 \hbar^2 \sum_{i<j} \left\{ \frac{\boldsymbol{S}_i \cdot \boldsymbol{S}_j}{r_{ij}^3} - \frac{3(\boldsymbol{S}_i \cdot \boldsymbol{r}_{ij})(\boldsymbol{S}_j \cdot \boldsymbol{r}_{ij})}{r_{ij}^5} \right\}, \qquad (7)$$

where $\gamma$ is the gyromagnetic ratio of the protons, and $\boldsymbol{S}_i$ and $\boldsymbol{S}_j$ represent their spin operators joined by the vector $\boldsymbol{r}_{ij}$, and $\gamma\hbar = g\beta$ (the nuclear $g$-factor times the Bohr magneton).

For a triplet state, eq 7 can be replaced by a phenomenological spin Hamiltonian described in terms of the total spin operator ($\boldsymbol{S} = \sum_i \boldsymbol{S}_i$):

$$\mathcal{H}_{SS} = \boldsymbol{S}^\mathrm{T}\boldsymbol{D}\boldsymbol{S} \qquad (8)$$

with $\boldsymbol{D}$ being the ZFS tensor. Choosing the PT direction throughout the base pair as the principal axis system (e.g., the $y$-direction) and the $z$-direction coincident with the DNA axis (see Figure 7a); such that $\boldsymbol{D}$ is diagonal, eq 8 reduces to

$$\mathcal{H}_{SS} = D_{SS}\left(S_z^2 - \frac{1}{3}\boldsymbol{S}^2\right) + E_{SS}(S_x^2 - S_y^2). \qquad (9)$$

The parameters $D_{SS}$ and $E_{SS}$ can be properly determined by the dipolar interaction and are, respectively, associated to the asymmetry of the spin distributions along the $z$-axis and in the $xy$-plane.

In the ZFS due to dipole-dipole interaction, the triplet eigenvectors are linear combinations of the eigenvectors of the $S_z$ operator, corresponding to the magnetic quantum numbers $M_S = +1, 0, -1$ (see refs 95 and 96):

$$\begin{cases} |T_x\rangle = \dfrac{1}{\sqrt{2}}(|\downarrow\downarrow\rangle - |\uparrow\uparrow\rangle) \\ |T_y\rangle = \dfrac{i}{\sqrt{2}}(|\downarrow\downarrow\rangle + |\uparrow\uparrow\rangle) \\ |T_z\rangle = \dfrac{1}{\sqrt{2}}(|\uparrow\downarrow\rangle + |\downarrow\uparrow\rangle) \end{cases} \qquad (10)$$

These states are maximally entangled (see eq 5) and satisfy $S_u|T_u\rangle = 0$, $S_u|T_v\rangle = i\hbar|T_w\rangle$ (cyclic), and $\boldsymbol{S}^2|T_u\rangle = 2\hbar^2|T_u\rangle$ ($u$, $v$, $w = x$, $y$, $z$). Thus, each triplet state $|T_u\rangle$ is an eigenstate of the operator $S_u$ with eigenvalue 0. As expected, there is no magnetic moment associated with any of the triplet states in zero magnetic field, since $\langle T_u|S_u|T_u\rangle = 0$ ($u = x, y, z$). However, it follows that magnetic dipole transitions are possible between any two ZFS sublevels (see Figure 7b).[97]

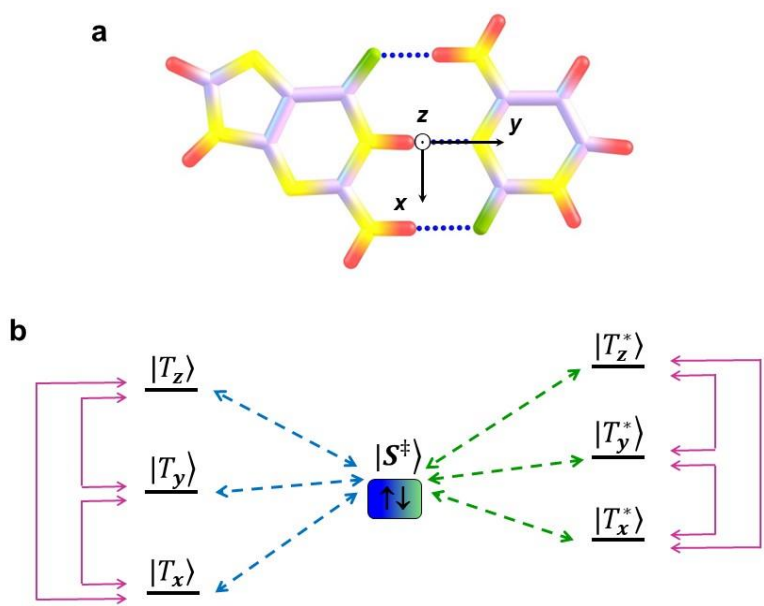

**Figure 7**. Schematic illustration of (a) the reference principal axis system and (b) the diagram for the singlet fission model from the TS in a DNA base pair.

**3.2.2. A singlet fission model for protons during the tautomerism.** The ZFS mechanism for the G*–C* pair is similar to that described by a similar Hamiltonian given by eq 9, differing only by a real constant. However, as the triplets are independent, i.e., $|T\rangle$ for G–C and $|T^*\rangle$ for G*–C*, and they are in a quantum superposition, in principle, there is no direct interaction between them. Indeed, within the WI2PM to deal with tautomersim both triplets only interconvert into each other, passing through a singlet state: $|T\rangle \leftrightharpoons |S^\ddagger\rangle \rightleftharpoons |T^*\rangle$, as depicted in Figure 6. In this sense, the tautomeric equilibrium may take place via a proton-hole interaction formed during PT. In fact, a model beyond the WI2PM would be viable by using two interacting proton-hole triplet pairs[97-99] centered in each nucleotide. In the simplest version of the two-proton two-hole picture, it is possible to consider the usual singlet fission mechanism,[83,86] for which one obtains 16 different combinations of four spins: 2 singlets, 9 triplets and 5 quintets.[100]

For a simpler picture, it is more convenient and sufficient to consider here the two-proton model consisting of two independent triplets. Furthermore, even though considering this simple model, it is possible to write a spin state for the protonic singlet fission that creates simultaneous triplet pairs, i.e.,

$$|S^\ddagger\rangle = \frac{1}{\sqrt{3}}\left(|T_x T_x^*\rangle + |T_y T_y^*\rangle + |T_z T_z^*\rangle\right) \quad (11)$$

Of course $|S^\ddagger\rangle$ is not an eigenstate of the ZFS Hamiltonian for the non-interacting triplet pairs, although these are the product states $|T_x T_x^*\rangle$, $|T_y T_y^*\rangle$, and $|T_z T_z^*\rangle$. In turn, these states acquire relative phases, evolving in time with frequencies defined by the energy sublevels (see Figure 7b). Thus, the WCQS is periodically in $|T\rangle$ for G–C or $|T^*\rangle$ for G*–C*, passing through $|S^\ddagger\rangle$ for G$^\ddagger$–C$^\ddagger$ via a rapid TTA mechanism. This spin-allowed mechanism increases the formation rate of G–C over G*–C* during the tautomeric equilibrium, but conserving the triplet entanglement.

**3.2.3. Indirect spin-spin coupling and the secular approximation.** It is also interesting to consider the role of the nuclear scalar indirect *J*-coupling for the two protons involved the tautomeric equilibrium of the base pairs.[101] In NMR spectroscopy, this is ideally suited for the construction of useful controlled gates for quantum computing.[102] Thus, the interaction of one H$^+$ is transmitted to the other through both covalent and H-bonds, formed and re-formed during the tautomeric process, mediated by the electronic coupling. Differently from the direct dipolar interaction, *J*-coupling does not average to zero under conditions of molecular tumbling, resulting in a perturbation of the NMR spectrum, which is independent of molecular orientation. Thus, while measuring dipolar interaction is most important in DNA crystals, *J*-coupling is important in folding double-stranded DNA.[103] This Hamiltonian (in frequency units) assumes the form

$$\mathcal{H}_J = 2\pi J \mathbf{S}_1 \cdot \mathbf{S}_2 \qquad (12)$$

Now, take into account an external magnetic field (B∥z) applied to an axis-oriented base pair (Figure 7), the total spin Hamiltonian for a coupled two-proton system can be written as

$$\mathcal{H}_{spin} = \mathcal{H}_Z + \mathcal{H}_J + \mathcal{H}_{SS} \qquad (13)$$

The Hamiltonian $\mathcal{H}_Z = \omega_0(S_{1z} + S_{2z})$ in eq 13 describes the Zeeman interaction, with $\omega_0 = g\beta B$ being the Larmor frequency, involving the nuclear *g*-factor, the Bohr magneton, and the magnetic field $B$. To simplify any further calculations, it is convenient to employ the secular approximation, dropping out all terms not along $B$, such that $\mathcal{H}_{SS} = d(3S_{1z}S_{2z} - \mathbf{S}_1 \cdot \mathbf{S}_2)$.

In the secular approximation, eq 13 can be simply rewritten[87] as

$$\mathcal{H}_{spin} = \mathcal{H}_A + \mathcal{H}_B \qquad (14)$$

where

$$\mathcal{H}_A = \omega_0(S_{1z} + S_{2z}) + 2\omega_A S_{1z}S_{2z} \text{ and } \mathcal{H}_B = \frac{\omega_B}{2}(S_1^+ S_2^- + S_1^- S_2^+) \qquad (15)$$

with $\omega_A = \pi J + d$ and $\omega_B = \pi J - d$.

In this approximation, $\mathcal{H}_A$ is diagonal while $\mathcal{H}_B$ possesses off-diagonal terms corresponding to the spin flip-flop transitions (see ref 87).

As, in general, a NMR-based quantum device comprises an ensemble of spins, it is convenient to calculate the density matrix, which can describe either a mixture or a pure state. For the problem of two interacting identical protons, the $(4 \times 4)$-density matrix $\rho^{(1,0)}$ in the singlet-triplet basis (eqs 5 and 6) can be represented as a direct sum of density matrices in two independent subspaces:[104]

$$\rho^{(1,0)} = (\rho^{(1)} \oplus \rho^{(0)}) \tag{16}$$

In this case, $\rho^{(1)}$ is the density matrix of a pseudo-qutrit and $\rho^{(0)}$ is the density matrix of a scalar singlet system. Thus, the time evolution of $\rho^{(1)}$ is determined from the Liouville equation, satisfying the requirement $\text{Tr}\rho^{(1)} = 1$,

$$\frac{\partial \rho^{(1)}}{\partial t} + i[\mathcal{H}^{(1)}, \rho^{(1)}] = 0 \tag{17}$$

where $\mathcal{H}^{(1)}$ should be determined in the triplet subspace. Additionally, the exact quantum dynamics of the two-spin system interacting with the environment can be obtained using the Markovian approximation.[105]

Considering only the triplet states for the two protons under the Hamiltonian given by eq 14 in the secular approximation, one can employ the triplet basis $\{|T_+\rangle = |\uparrow\uparrow\rangle, |T_0\rangle = |T_z\rangle, |T_-\rangle = |\downarrow\downarrow\rangle\}$ (the Zeeman basis) to obtain its diagonal matrix representation

$$\mathcal{H}^{(1)}_{spin} = \frac{1}{2}\begin{pmatrix} 2\omega_0 + \pi J + d & 0 & 0 \\ 0 & \pi J - 2d & 0 \\ 0 & 0 & -2\omega_0 + \pi J + d \end{pmatrix} \tag{18}$$

It is clear that the energy levels of this pseudo-qutrit system depend on the relative orientations of the two dipoles with respect to the external magnetic field.[104] However, in the absence of an external magnetic field, the energy levels do not depend on the relative orientation of the dipoles. Indeed, they are non-degenerate because of the ZFS interaction, and possess maximally entangled eigenvectors (see eq 10).

# 4. A QUANTUM-COMPUTING APPROACH

## 4.1. Qubit Representations and Intrinsic Gates in DNA Base Pairs

It is now important to examine in which conditions DNA base pairs can work as quantum processing units (QPUs)[106] driven by qubit or pseudo-qutrit operations.[107] First, from the WI2PM, proton spins might be less vulnerable to decoherence than their spatial degree of freedom. Second, the prototropic ground states are protected by symmetry, which avoids arbitrary mixing spin up and down states in the proton eigenstates. Third, the spin coherence time of the triplet states must be sufficiently longer than single- and two-qubit operation times, which can be controlled, for example, by switching magnetic fields and gates.[108] Four, magnetic hyperfine interactions between electrons and protons are small. Otherwise, these interactions can be suppressed by applying an external magnetic field or through the Overhause effect.[108,109]

Take into account these considerations, it appears that DNA base pairs, in principle, satisfy the necessary and sufficient conditions for quantum computing.[106] Furthermore, the thermal stability of the tautomerism allows quantum computation even at the room temperature, since these nonlocal states are macroscopically entangled in DNA.[64] In this sense, encoding quantum information in each base pair of DNA can be realized from, at least, two-qubit operations. The single-proton qubits can be defined as $|\uparrow\rangle \equiv |0\rangle$ and $|\downarrow\rangle \equiv |1\rangle$, so that it is possible writing how intrinsic one-qubit quantum gates (e.g.,[110] $I, X, Y, Z, S, T$, or $H$) act on single-proton spin states. For example, the Pauli $X$ gate or NOT gate turns $|0\rangle$ into $|1\rangle$ and $|1\rangle$ into $|0\rangle$.

In this context, using the Zeeman basis, a general triplet state[64] for a canonical base pair can be described, before a measurement, by the general superposition:

$$|T\rangle = \frac{1}{\sqrt{3}}\left[|00\rangle + |11\rangle + \frac{1}{\sqrt{2}}(|01\rangle + |10\rangle)\right] \quad (20)$$

The action of an intrinsic two-qubit quantum gate on $|T\rangle$ leads to the tautomeric triplet state:

$$|T^*\rangle = \frac{1}{\sqrt{3}}\left[|11\rangle + |00\rangle + \frac{1}{\sqrt{2}}(|10\rangle + |01\rangle)\right] \qquad (21)$$

to within a global phase factor. From single-qubit operations, a tensorial representation of Pauli *X* gates can transform the CQS into TQS, and vice-versa, for *N* successive processes. Hence, intrinsically, the tautomeric equilibrium results by action of the Pauli *X* gates. For practical purposes, any probability-preserving linear map acting on the spin states of the base pairs can be a valid intrinsic quantum gate.

Notwithstanding, the underlying quantum machinery associated to the tautomeric equilibrium of DNA base pairs is not so simple and requires a better description. Going to a little further, it is chosen a four-qubit representation of the form $|i\bar{j}\bar{k}l\rangle$, where the leftmost qubit $|i\rangle$ corresponds to an occupied state with spin *i* in G, and the rightmost qubit $|l\rangle$ corresponds to an occupied state with spin *l* in C. The qubits $|\bar{j}\rangle$ and $|\bar{k}\rangle$ appearing in the middle of the four-qubit correspond to unoccupied spin states and follow a rule similar to that used for the occupied states, being $|\bar{j}\rangle$ in G and $|\bar{k}\rangle$ in C. Thus, the triplet state components in G–C is represented by (using the leftmost qubit in the bottom and the rightmost in the top):

$$_G\begin{bmatrix} & \uparrow \\ \uparrow & \end{bmatrix}_C \Leftrightarrow |0\bar{1}\bar{1}0\rangle, \quad _G\begin{bmatrix} & \downarrow \\ \downarrow & \end{bmatrix}_C \Leftrightarrow |1\bar{0}\bar{0}1\rangle,$$

and

$$_G\begin{bmatrix} & \downarrow \\ \uparrow & \end{bmatrix}_C + _G\begin{bmatrix} & \uparrow \\ \downarrow & \end{bmatrix}_C \Leftrightarrow \frac{1}{\sqrt{2}}(|0\bar{1}\bar{0}1\rangle + |1\bar{0}\bar{1}0\rangle).$$

These representations encompasses the tautomeric G*–C* basis pair by flipping the spins and addressing the occupied/unoccupied states. In this way, the tautomeric pair conserves the triplet-entangled state.

As an example, taking into account the ADPT mechanism via a stepwise transfer from the canonical structure, one can represent it as a two-step process (i) in terms of the zwitterionic single PT[63] from G to C:

$$_G^-[\cdots\rightarrow \uparrow\downarrow]_C^+ \Leftrightarrow |\bar{0}\bar{1}10\rangle, \quad _G^-[\cdots\rightarrow \downarrow\uparrow]_C^+ \Leftrightarrow |\bar{1}\bar{0}01\rangle,$$

and

$$_G^-[\cdots\rightarrow \downarrow\downarrow]_C^+ + _G^-[\cdots\rightarrow \uparrow\uparrow]_C^+ \Leftrightarrow \frac{1}{\sqrt{2}}(|\bar{0}\bar{1}11\rangle + |\bar{1}\bar{0}00\rangle).$$

Alternatively, (ii) in terms of the zwitterionic single PT[63] from C to G:

$$_G^+[\downarrow\uparrow \leftarrow\cdots]_C^- \Leftrightarrow |01\bar{1}\bar{0}\rangle, \quad _G^+[\uparrow\downarrow \leftarrow\cdots]_C^- \Leftrightarrow |10\bar{0}\bar{1}\rangle,$$

and

$$_G^+[\uparrow\uparrow \leftarrow\cdots]_C^- + _G^+[\downarrow\downarrow \leftarrow\cdots]_C^- \Leftrightarrow \frac{1}{\sqrt{2}}(|00\bar{0}\bar{1}\rangle + |11\bar{1}\bar{0}\rangle).$$

Using this four-qubit basis to obtain the triplet tautomeric state, it is necessary to apply proper operators. For example, starting with the state $|0\bar{1}\bar{1}0\rangle$ of G–C and following the two-step process (i), one obtains $|\bar{0}11\bar{0}\rangle$ of G*–C* by applying the quantum operations:

$$C \begin{Bmatrix} |0\rangle \xrightarrow{I} |0\rangle \xrightarrow{a_0} |\bar{0}\rangle \\ |\bar{1}\rangle \xrightarrow{a_{\bar{1}}^\dagger} |1\rangle \xrightarrow{I} |1\rangle \end{Bmatrix} C^*$$

$$G \begin{Bmatrix} |\bar{1}\rangle \xrightarrow{\bar{I}} |\bar{1}\rangle \xrightarrow{a_{\bar{1}}^\dagger} |1\rangle \\ |0\rangle \xrightarrow{a_0} |\bar{0}\rangle \xrightarrow{\bar{I}} |\bar{0}\rangle \end{Bmatrix} G^*$$

By following the two-step process (ii) one applies the quantum operations:

$$C \begin{Bmatrix} |0\rangle \xrightarrow{a_0} |\bar{0}\rangle \xrightarrow{\bar{I}} |\bar{0}\rangle \\ |\bar{1}\rangle \xrightarrow{\bar{I}} |\bar{1}\rangle \xrightarrow{a_{\bar{1}}^\dagger} |1\rangle \end{Bmatrix} C^*$$

$$G \begin{Bmatrix} |\bar{1}\rangle \xrightarrow{a_{\bar{1}}^\dagger} |1\rangle \xrightarrow{I} |1\rangle \\ |0\rangle \xrightarrow{I} |0\rangle \xrightarrow{a_0} |\bar{0}\rangle \end{Bmatrix} G^*$$

The one-qubit operations are defined now by the identity gates ($I$ and $\bar{I}$) acting on the occupied and unoccupied qubits, respectively. Additionally, $a_i^\dagger$ and $a_j$ are creation and annihilation operators, respectively, acting on spins $i$ and $j$ at occupied or virtual states.

Notice that in this four-qubit basis considering the formation of zwitterionic states, the tautomeric equilibrium is not simply defined in terms of Pauli $X$ gates. Moreover, it needs the action of creation and annihilation operators. Thus, an exciting possibility is to consider the tautomeric two-proton dynamics in DNA base pairs in a specific quantum device by applying appropriate fermion-qubit transformations.[111,112] For instance, returning to the Hamiltonian given by eq 2, it can be calculated through a universal quantum simulator.[113] In this way, the fermion-qubit transformations convert the occupation number basis to qubit states, and describe the fermionic creation/annihilation operators in terms of Pauli gates acting on qubits.

## 4.2. Toward a Quantum Computer Simulation of a DNA Base Pair

Before proposing DNA as a perfect quantum computer,[114] it is timely to explain how to simulate the model in a feasible quantum device.[115] For simplicity, considering a rigid base pair, only the two terms of eq 2, i.e., $\mathcal{H}_Q^{H^+}$ and $\mathcal{H}_{coup}^{XY-H^+}$, need to be considered in the 'protonic' Hamiltonian, since $\mathcal{H}_{BO}^{XY}$ is clamped. Thus, using the single-proton space functions $\{\psi_i\}$ as an initial minimal basis, one obtains the second quantized representation for the confined two-proton model

$$\mathcal{H}_{prot} = \sum_{pq=1}^{M} h_{pq} a_p^\dagger a_q + \frac{1}{2} \sum_{pqrs=1}^{M} h_{pqrs} a_p^\dagger a_q^\dagger a_r a_s \quad (23)$$

with the creation/annihilation operators satisfying the fermionic anti-commutation relations $\{a_j, a_k\} = \{a_j^\dagger, a_k^\dagger\} = 0, \{a_j, a_k^\dagger\} = \delta_{jk}$.

The one-proton integrals in eq 23 assume the form (in atomic units)

$$h_{pq} = \int d\mathbf{r}\, \psi_p^*(\mathbf{r}) \left( \frac{-\nabla^2}{2m_{H^+}} + \mathcal{H}_{coup}^{XY-H^+} \right) \psi_q(\mathbf{r}) \quad (24)$$

while the two-proton integrals become

$$h_{pqrs} = \iint \frac{d\mathbf{r}_1 d\mathbf{r}_2\, \psi_p^*(\mathbf{r}_1) \psi_q^*(\mathbf{r}_2) \psi_r(\mathbf{r}_1) \psi_s(\mathbf{r}_2)}{|\mathbf{r}_1 - \mathbf{r}_2|} \quad (25)$$

In eqs 24 and 25, $\psi_i$ functions are single-proton space functions of the X–Y base pair as proposed in 3.1.2., which might be multiplied by the appropriate single-proton spin state.

It is important to notice that the unknowing functions $\{\psi_i\}$ can be determined, for example, self-consistently using a general approach for learning the long-range response of electrostatics in neuronal network potentials.[116] Then, it is possible, in principle, to determine indirectly these integrals. Recently, Shee *et al.*[117] have proposed a hybrid quantum chemistry-quantum computation method to simulate a preferred tautomeric state prediction, which can also benefit the Hamiltonian partitioning such as proposed by the WI2PM. Thus, if one implements such an approach for two interacting protons, $\mathcal{H}_{prot}$ can be determined for any specific DNA base pair.

From this formalism, the Hamiltonian given by eq 23 with $M = 2$ can be mapped onto two qubits using, for example, the Bravyi-Kitaev transformation,[112] leading to a scalable qubit Hamiltonian[113] of the form

$$\mathcal{H}_{qb} = g_0 I + g_1 Z_0 + g_2 Z_1 + g_3 Z_0 Z_1 + g_4 Y_0 Y_1 + g_5 X_0 X_1 \quad (26)$$

With this scheme, it is possible to implement eq 26 in a quantum device. The scalars $\{g_i\}$ are real numbers determined by equilibrium structural parameters of the base pairs, and $\{I, X_i, Y_i, Z_i\}$ are the identity and Pauli gates acting on the $i$-th qubit. For this case, one can employ the variational quantum eigensolver (VQE) method[118] to find both the canonical and tautomeric ground states of the base pairs, as well as their separation during DNA replication.

It is not the purpose of this paper to simulate or deal with a specific algorithm to obtain the quantum states of protons in DNA base pairs. However, one of the aims is to make clear that, once such states there exist (as demonstrated in Section 3), the base pairs can be used for quantum computing. Clearly, several available algorithms, including VQE, to simulate two-fermion systems[119-122] could be utilized in this case. Furthermore, a parametrized quantum circuit[123] could be constructed to generate the probability densities of the protons in the tautomeric equilibrium. In this sense, CQS and TQS can be established and defined in a proper quantum superposition.

**4.3. Quantum Logic Gates and Controlled Operations in DNA Base Pairs**

The possibility of using DNA molecules for a quantum computer have been explored in the context of the oscillatory resonant quantum states of correlated electron–hole pairs, where the central H-bond in G–C or A–T may work as an ideal Josephson-junction qubit.[114] Herein, an alternative and feasible model is to explore the entangled protons in each base pair as protected elementary units of information for an NMR-based quantum computer.[102,124] Then, in principle, a viable method is to employ the spins of the protons, directly involved in the prototropic tautomerism, to implement quantum gates using appropriate radiofrequency pulses in NMR spectroscopy.[87,125,126]

In this type of DNA-based NMRQC, single-qubit gates rotations are implemented by means of a Ramsey sequence and can be tuned using a closed-loop optimization with a randomized benchmark.[127] Additionally, a qubit state measurement[128] can be performed in a dispersive readout with NMR spectroscopy. For example, in crystallized DNA samples, the coherent dipolar spin-exchange interactions can be experimentally investigated and modeled by using the ZFS Hamiltonian (eq 8) within the secular approximation (eq 15). Thus, as in related studies,[129-131] Ramsey pulses can be employed to retain long coherence times and produce specific qubit superposition states.[132]

As a hypothetical example, starting with the CQS prepared in $|T_+\rangle = |\uparrow\uparrow\rangle$ ($M = +1$), it is possible to create maximally entangled states using single and two-qubit gates within the ZFS approximation (see Figure 8). Thus, from a sequence for Bell states preparation, one converts a base pair quantum state, initially prepared in $|\uparrow\uparrow\rangle$, into $|T_x\rangle$ and $|T_y\rangle$ (see eq 10). Otherwise, if the base pair is prepared in a Zeeman state $|T_0\rangle = |T_z\rangle$ ($M = 0$), its time evolution under $\mathcal{H}_B$ (eq 15) leads to $e^{\frac{-i\omega_B t}{4}} \frac{1}{\sqrt{2}}(|\downarrow\uparrow\rangle + |\uparrow\downarrow\rangle)$, oscillating with frequency $\omega_B$. Also, during the time evolution, a relative phase accumulates between

$|T_z\rangle$ and $|T_x\rangle$, and between $|T_z\rangle$ and $|T_y\rangle$. To read out a final qubit state, a controlled pulse can project the CQS or TQS into $|T_+\rangle = |\uparrow\uparrow\rangle$ ($M = +1$) or $|T_-\rangle = |\downarrow\downarrow\rangle$ ($M = -1$).

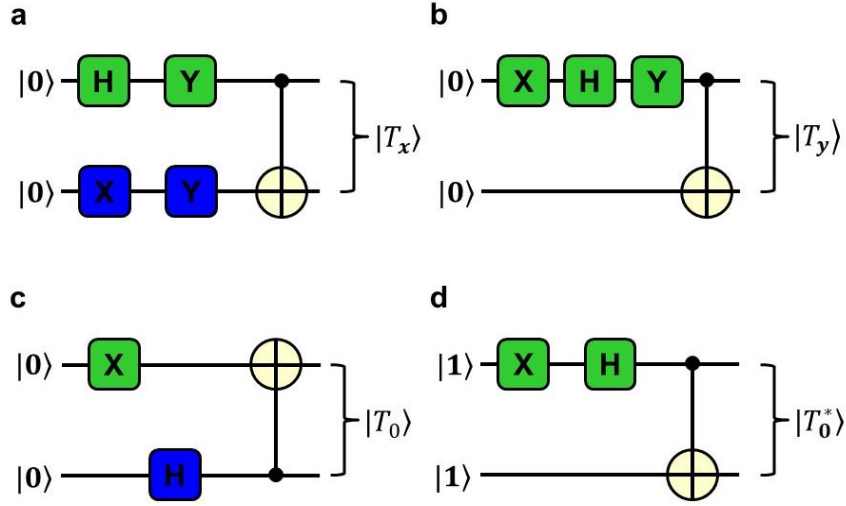

**Figure 8**. Quantum circuits for Bell states preparation of a DNA base pair with the ZFS. The triplet state component in $|00\rangle = |\uparrow\uparrow\rangle$ going to $|T_x\rangle$ (a), $|T_y\rangle$ (b), and $|T_z\rangle$ (c) using proper quantum gates. The TQS in $|11\rangle = |\downarrow\downarrow\rangle$ going to $|T_z^*\rangle$ (d). These states are prepared by applying single and two-qubit quantum gates; i.e., the Pauli *X* and *Y* gates, the Hadamard gate *H*, and proper CNOT gates (see ref 110). The blue colored gate acts on G, while the green one acts on C in this example.

Clearly, the dipolar spin-exchange interaction can be used to create a two-qubit gate, leading to entanglement in a base pair or between CQS and TQS. Considering the period of the spin-exchange oscillation, the system evolves into a maximally entangled state or a Bell state. Following Holland *et al.*,[131] by applying a Ramsey sequence that leaves $\mathcal{H}_B$ invariant, the spin-exchange interaction is preserved.[133] Thus, starting with the state $|\uparrow\uparrow\rangle$ one can obtain, for example, a time dependent superposition of the form $e^{\frac{-i\omega_B t}{4}}[i\sin(\frac{\omega_B t}{4})|\uparrow\uparrow\rangle - \cos(\frac{\omega_B t}{4})|\downarrow\downarrow\rangle]$, with oscillating probability $P_{\downarrow\downarrow}$. A sequence like that can be used to demonstrate the influence of the coherent spin-exchange interaction in a base pair.

Within this theoretical approach, controlling proton spin qubits magnetically offers the potential to reduce decoherence and increase the working temperature of an NMRQC. In this scenario, the spin-exchange interaction and spin dynamics, mediated by a deterministic triplet state in a base pair, effectively constitute a quantum-circuit building block. This reinforces the possibility of using DNA base pairs to implement controlled gates and to perform quantum computations, albeit it deserves further attention.

## 5. OUTLOOK

This paper demonstrates, as a proof-of-principle, that the underlying PT mechanism in DNA base pairs is deterministically entangled, which is crucial for quantum computing and information processing. From quantum mechanical principles, the two protons involved in the prtotropic tautomerism form interchangeable triplet states between the CQS and TQS protected by symmetry. Although these quantum states are not equally populated ($|a(T)|^2 \gg |b(T)|^2$), they are supposed to be in a coherent Watson-Crick superposition state. The dipolar spin-exchange interaction can convert the base pairs in elementary two-qubit QPUs. In this sense, quantum information of a DNA molecule is encrypted in its proton density and may be properly retrieved. This is only a basic idea behind the development of a DNA-based quantum computer.

Using this theoretical approach, two groundbreaking concepts emerge for the development of cut-edging computational technologies. On the one hand, DNA computing, that employs DNA molecules to carry out massively parallel calculations, potentially including large data storage and high performance computing. On the other hand, quantum computing, that processes information by employing superposition states and entangled qubits, and boosts computing power. These technologies, combined in one unique platform, DNA, are capable to envisage challenging issues that have long exceeded traditional silicon-based computing. However, several limitations need to be overcome before DNA can be proposed as a perfect quantum computer.

The main challenge is to know if a coherent large-scale order emerges from the immense number of base pairs contained in a DNA segment. Furthermore, techniques for DNA manipulation and computation face difficulties in reaching the scalability to treat complexes problems. DNA molecules are susceptible to degradation and influence of external factors, such as environment and experimental conditions, which lead to the loss of quantum coherence and collapse of the superposition states. Therefore, developing efficient and robust error-correction coding is fundamental for achieving reliable and scalable DNA-based quantum computers. It is also essential to carry out experiments in order to establish whether the spatial coherence of protons in a base pair survives at high temperatures. Only in this way, the PT with spin dynamics must be thought of as a large-scale coherent process for room temperature quantum computing with macroscopic control.

Regarding the impact of quantum information in the genetic code, the proposed model is still far from investigating, for instance, if DNA is held together by quantum entanglement. Perhaps, DNA base pairs could be a natural subsystem to manipulate quantum information using triplet states as two-qubit operations, pseudo-qutrits or, even, multiple qubits, which offer a way to perform more complex quantum computing and cryptography. Indeed, with the insights gained from the model proposed here, it appears that quantum information of a DNA molecule is encoded in the proton density inside its base pairs. However, new theoretical, computational, and experimental efforts need to be developed in order to get a proof-of-concept for a DNA-based quantum machine of this kind.

## ASSOCIATED CONTENT

**Supporting Information**

The Supporting Information is available free of charge at https://

## AUTHOR INFORMATION

**Corresponding Author**

**Roberto Rivelino** – *Instituto de Física, Universidade Federal da Bahia, 40210-340 Salvador, Bahia, Brazil;* https://orcid.org/0000-0003-2679-1640; Email: rivelino@ufba.br

## ACKNOWLEDGMENTS

This work was partially financed by the Brazilian agencies Coordenação de Aperfeiçoamento de Pessoal de Nível Superior (CAPES), Finance Code 001, Conselho Nacional de Desenvolvimento Científico e Tecnológico (CNPq), Grant No. 307826/2018-0, and Fundação de Amparo à Pesquisa do Estado da Bahia (FAPESB), Grant No 12/2023. The author also thanks INCT-FCx (CNPq) for the institutional facilities.

## REFERENCES

(1) Annunziato, A. DNA Packaging: Nucleosomes and Chromatin. *Nat. Edu.* **2008**, *1*, 26.

(2) Vignuzzi, M.; López, C. B. Defective viral genomes are key drivers of the virus–host interaction. *Nat. Microbiol.* **2019**, *4*, 1075–1087.

(3) Wessner, D. R. The Origins of Viruses. *Nat. Edu.* **2010**, *3*, 37.

(4) Portin, P.; Wilkins, A. The Evolving Definition of the Term "Gene", *Genetics* **2017**, *205*, 1353–1364.

(5) Shendure, J.; Balasubramanian, S.; Church, G. M.; Gilbert, W.; Rogers, J.; Schloss, J. A.; Waterston, R. H. DNA sequencing at 40: past, present and future. *Nature* **2017**, *550*, 345–353.

(6) Tabatabaei, S. K.; Pham, B.; Pan, C.; Liu, J.; Chandak, S.; Shorkey, S. A.; Hernandez, A. G.; Aksimentiev, A.; Chen, M.; Schroeder, C. M.; Milenkovic, O. Expanding the Molecular Alphabet of DNA-Based Data Storage Systems with Neural Network Nanopore Readout Processing. *Nano Lett*. **2022**, *22*, 1905–1914

(7) Betts, H. C.; Puttick, M. N.; Clark, J. W.; Williams, T. A.; Donoghue, P. C. J.; Pisani, D. Integrated genomic and fossil evidence illuminates life's early evolution and eukaryote origin. *Nat. Ecol. Evol*. **2018**, *2*, 1556–1562.

(8) Schopf, J. W.; Kitajima, K.; Spicuzza, M. J.; Kudryavtsev, A. B.; Valley, J. W. SIMS analyses of the oldest known assemblage of microfossils document their taxon-correlated carbon isotope compositions. *Proc. Natl. Acad. Sci. U.S.A*. **2018**, *115*, 53–58.

(9) Dalla Pozza, M.; Abdullrahman, A.; Cardin, C. J.; Gasser, G.; Hall, J. P. Three's a crowd - stabilisation, structure, and applications of DNA triplexes. *Chem. Sci*. **2022**, *13*, 10193-10215.

(10) Varshney, D.; Spiegel, J.; Zyner, K.; Tannahill, D.; Balasubramanian, S. The regulation and functions of DNA and RNA G-quadruplexes. *Nat. Rev. Mol. Cell. Biol*. **2020**, *21*, 459–474.

(11) Matta, C. F; Castillo, N.; Boyd, R. J. Extended weak bonding interactions in DNA: pi-stacking (base-base), base-backbone, and backbone-backbone interactions. *J. Phys. Chem. B* **2006**, *110*, 563–578.

(12) Poater, J.; Swart, M.; Bickelhaupt, F. M.; Guerra, C. F. B-DNA structure and stability: the role of hydrogen bonding, π–π stacking interactions, twist-angle, and solvation. *Org. Biomol. Chem.* **2014**, *12*, 4691–4700.

(13) Watson, J.; Crick, F. Molecular Structure of Nucleic Acids: A Structure for Deoxyribose Nucleic Acid. *Nature* **1953**, *171*, 737–738.

(14) Nieuwland, C.; Hamlin, T. A.; Guerra, C. F.; Barone, G.; Bickelhaupt, F. M. B-DNA Structure and Stability: The Role of Nucleotide Composition and Order. *ChemistryOpen* **2022**, *11*, e202100231.

(15) Ping, Z.; Ma, D.; Huang, X.; Chen, S.; Liu, L.; Guo, F.; Zhu, S. J.; Shen, Y. Carbon-based archiving: current progress and future prospects of DNA-based data storage, *GigaScience* **2019**, 8, giz075.

(16) Yang, S., Bögels, B. W. A., Wang, F.; Xu, C.; Dou, H.; Mann, S.; Fan, C.; de Greef, T. F. A. DNA as a universal chemical substrate for computing and data storage. *Nat. Rev. Chem.* **2024**, *8*, 179–194.

(17) Lin, K. N.; Volkel, K.; Cao, C.; Hook, P. W.; Polak, R. E.; Clark, A. S.; San Miguel, A.; Timp, W.; Tuck, J. M.; Velev, O. D.; Keung, A. J. A primordial DNA store and compute engine. *Nat. Nanotechnol.* **2024**, doi: 10.1038/s41565-024-01771-6.

(18) Church, G. M.; Gao, Y.; Kosuri, S. Next-generation digital information storage in DNA. *Science* **2012**, 337, 1628.

(19) Takahashi, C. N.; Nguyen, B. H.; Strauss, K.; Ceze, L. Demonstration of End-to-End Automation of DNA Data Storage. *Sci. Rep.* **2019**, 9, 4998.

(20) Zhang, Y.; Kong, L.; Wang, F.; Li, B.; Ma, C.; Chen, D.; Liu, K.; Fan, C.; Zhang, H. Information stored in nanoscale: Encoding data in a single DNA strand with Base64. *Nano Today* **2020**, *33*, 100871.

(21) Ding, L.; Wu, S.; Hou, Z.; Li, A.; Xu, Y.; Feng, H.; Pan, W.; Ruan, J. Improving error-correcting capability in DNA digital storage via soft-decision decoding. *Natl. Sci. Rev.* **2023**, *11*, nwad229.

(22) Chung, G.; Oh, H. B.; Lee, D. Tautomerism and isomerism of guanine–cytosine DNA base pair: Ab initio and density functional theory approaches. *J. Mol. Struct. Theochem* **2005**, *730*, 241-249.

(23) Person, W. B.; Szczepaniak, K.; Szczesniak, M.; Kwiatkowski, J. S.; Hernandez, L.; Czerminski, R. Tautomerism of nucleic acid bases and the effect of molecular interactions on tautomeric equilibria. *J. Mol. Struct.* **1989**, *194*, 239-258.

(24) Colominas, C.; Luque, F. J.; Orozco, M. Tautomerism and Protonation of Guanine and Cytosine. Implications in the Formation of Hydrogen-Bonded Complexes. *J. Am. Chem. Soc.* **1996**, *118*, 6811–6821.

(25) Angiolari, F.; Huppert, S.; Pietrucci, F.; Spezia, R. Environmental and Nuclear Quantum Effects on Double Proton Transfer in the Guanine–Cytosine Base Pair. *J. Phys. Chem. Lett.* **2023**, *14*, 5102–5108.

(26) Lin, Y.; Wang, H.; Gao, S.; Li, R.; Schaefer, III, H. F. Hydrogen-Bonded Double-Proton Transfer in Five Guanine−Cytosine Base Pairs after Hydrogen Atom Addition. *J. Phys. Chem. B* **2012**, *116*, 8908−8915.

(27) Chen, H.-Y.; Kao, C.-L.; Hsu, S. C. N. Proton Transfer in Guanine-Cytosine Radical Anion Embedded in B-Form DNA. *J. Am. Chem. Soc.* **2009**, *131*, 15930–15938.

(28) Jacquemin, D.; Zúñiga, J.; Requena, A.; Céron-Carrasco, J. P. Assessing the Importance of Proton Transfer Reactions in DNA. *Acc. Chem. Res.* **2014**, *47*, 2467−2474.

(29) Löwdin, P.-O. Proton Tunneling in DNA and its Biological Implications *Rev. Mod. Phys.* **1963**, *35*, 724 – 732.

(30) Kryachko, E. S. The Origin of Spontaneous Point Mutations in DNA via Löwdin Mechanism of Proton Tunneling in DNA Base Pairs: Cure with Covalent Base Pairing. *Int. J. Quantum Chem.* **2002**, *90*, 910−923.

(31) Xiao, S.; Wang, L.; Liu, Y.; Lin, X.; Liang, H. Theoretical investigation of the proton transfer mechanism in guanine-cytosine and adenine-thymine base pairs. *J. Chem. Phys.* **2012**, *137*, 195101.

(32) Sauri, V.; Gobbo, J. P.; Serrano-Pérez, J. J.; Lundberg, M.; Coto, P. B.; Serrano-Andrés, L.; Borin, A. C.; Lindh, R.; Merchán, M.; Roca-Sanjuán, D. Proton/Hydrogen

Transfer Mechanisms in the Guanine−Cytosine Base Pair: Photostability and Tautomerism. *J. Chem. Theory Comput.* **2013**, *9*, 481–496.

(33) Umesaki, K.; Odai, K. A Kinetic Approach to Double Proton Transfer in Watson–Crick DNA Base Pairs. *J. Phys. Chem. B* **2020**, *124*, 1715-1722.

(34) Xue, J., Guo, X., Wang, X.; Xiao, Y. Density functional theory studies on cytosine analogues for inducing double-proton transfer with guanine. *Sci. Rep.* **2020**, *10*, 9671.

(35) Topal, M. D.; Fresco, J. R. Complementary base pairing and the origin of substitution mutations. *Nature* **1976**, *263*, 285–289.

(36) Florián, J.; Leszczyńsk, J. Spontaneous DNA Mutations Induced by Proton Transfer in the Guanine·Cytosine Base Pairs: An Energetic Perspective. *J. Am. Chem. Soc.* **1996**, *118*, 3010–3017.

(37) Kumar, A.; Sevilla, M. D. Influence of Hydration on Proton Transfer in the Guanine−Cytosine Radical Cation ($G^{•+}$−C) Base Pair: A Density Functional Theory Study. *J. Phys. Chem. B* **2009**, *113*, 11359–11361.

(38) Pérez, A.; Tuckerman, M. E.; Hjalmarson, H. P.; von Lilienfeld, O. A. Enol Tautomers of Watson−Crick Base Pair Models Are Metastable Because of Nuclear Quantum Effects. *J. Am. Chem. Soc.* **2010**, *132*, 11510–11515.

(39) Lin, Y.; Wang, H.; Gao, S.; Schaefer III, H. F. Hydrogen-Bonded Proton Transfer in the Protonated Guanine-Cytosine $(GC^+H)^+$ Base Pair. *J. Phys. Chem. B* **2011**, *115*, 11746–11756.

(40) Villani, G. Coupling Between Hydrogen Atoms Transfer and Stacking Interaction in Adenine-Thymine/Guanine-Cytosine Complexes: A Theoretical Study. *J. Phys. Chem. B* **2014**, *118*, 5439–5452.

(41) Freitas, R. R. Q.; Rivelino, R.; de Brito Mota, F.; Gueorguiev, G. K.; de Castilho, C. M. C. Energy Barrier Reduction for the Double Proton-Transfer Reaction in Guanine–Cytosine DNA Base Pair on a Gold Surface. *J. Phys. Chem. C* **2015**, *119*, 15735–15741.

(42) Das, S.; Nam, K.; Major, D. T. Rapid Convergence of Energy and Free Energy Profiles with Quantum Mechanical Size in Quantum Mechanical–Molecular Mechanical Simulations of Proton Transfer in DNA. *J. Chem. Theory Comput.* **2018**, *14*, 1695–1705.

(43) Soler-Polo, D.; Mendieta-Moreno, J. I.; Trabada, D. G.; Mendieta, J.; Ortega, J. Proton Transfer in Guanine-Cytosine Base Pairs in B-DNA. *J. Chem. Theory Comput.* **2019**, *15*, 6984–6991.

(44) Gheorghiu, A.; Coveney, P. V.; Arabi, A. A. The influence of base pair tautomerism on single point mutations in aqueous DNA. *Interface Focus* **2020**, *10*, 20190120.

(45) Li, P.; Rangadurai, A.; Al-Hashimi, H. M.; Hammes-Schiffer, S. Environmental Effects on Guanine-Thymine Mispair Tautomerization Explored with Quantum Mechanical/Molecular Mechanical Free Energy Simulations. *J. Am. Chem. Soc.* **2020**, *142*, 11183-11191.

(46) Slocombe, L.; Winokan, M.; Al-Khalili, J.; Sacchi, M. Proton transfer during DNA strand separation as a source of mutagenic guanine-cytosine tautomers. *Commun. Chem.* **2022**, *5*, 144.

(47) Pohl, R.; Socha, O.; Slavíček, P.; Šála, M.; Hodgkinson, P.; Dračínský, M. Proton transfer in guanine–cytosine base pair analogues studied by NMR spectroscopy and PIMD simulations. *Faraday Discuss*. **2018**, *212*, 331–334.

(48) Slocombe, L.; Sacchi, M.; Al-Khalili, J. An open quantum systems approach to proton tunnelling in DNA. *Commun. Phys*. **2022**, *5*, 109.

(49) Eyring, H. The Activated Complex in Chemical Reactions. *J. Chem. Phys*. **1935**, *3*, 107–115.

(50) Truhlar, D. G.; Garrett, B. C.; Klippenstein, S. J. Current Status of Transition-State Theory. *J. Phys. Chem.* **1996**, *100*, 12771–12800.

(51) Slocombe, L.; Al-Khalilib, J. S.; Sacchi, M. Quantum and classical effects in DNA point mutations: Watson–Crick tautomerism in AT and GC base pairs. *Phys. Chem. Chem. Phys*. **2021**, *23*, 4141–4150.

(52) King, B.; Winokan, M.; Stevenson, P.; Al-Khalili, J.; Louie Slocombe, L.; Sacchi, M. Tautomerisation Mechanisms in the Adenine-Thymine Nucleobase Pair during DNA Strand Separation. *J. Phys. Chem. B* **2023**, *127*, 4220–4228.

(53) Slocombe, L.; Winokan, M.; Al-Khalili, J.; Sacchi, M. Quantum Tunnelling Effects in the Guanine-Thymine Wobble Misincorporation via Tautomerism. *J. Phys. Chem. Lett*. **2023**, *14*, 9–15.

(54) Brunet, A.; Salomé, L.; Rousseau, P.; Destainville, N.; Manghi, M.; Tardin, C. How does temperature impact the conformation of single DNA molecules below melting temperature? *Nucleic Acids Res.* **2018**, *46*, 2074–2081.

(55) Shakked, Z.; Guerstein-Guzikevich, G.; Eisenstein, M.; Frolow, F.; Rabinovich. D. The conformation of the DNA double helix in the crystal is dependent on its environment. *Nature* **1989**, *342*, 456–460.

(56) Zhang, C.; Xie, L.; Ding, Y.; Sun, Q.; Xu, W. Real-Space Evidence of Rare Guanine Tautomer Induced by Water. *ACS Nano* **2016**, *10*, 3776–3782.

(57) Fang, Y.-G.; Valverde, D.; Mai, S.; Canuto, S.; Borin, A. C.; Cui, G.; González, L. Excited-State Properties and Relaxation Pathways of Selenium-Substituted Guanine Nucleobase in Aqueous Solution and DNA Duplex. *J. Phys. Chem. B* **2021**, *125*, 1778–1789.

(58) Gheorghiu, A.; Coveney, P. V.; Arabi, A. A. The influence of external electric fields on proton transfer tautomerism in the guanine–cytosine base pair. *Phys. Chem. Chem. Phys*. **2021**, *23*, 6252–6265.

(59) Cerón-Carrasco, J. P.; Jacquemin, D. Electric-field induced mutation of DNA: a theoretical investigation of the GC base pair. *Phys. Chem. Chem. Phys.* **2013**, *15*, 4548-4553.

(60) Driessen, R. P.; C.; Sitters, G.; Laurens, N.; Moolenaar, G. F.; Wuite, G. J. L.; Goosen, H.; Dame, R. Th., Effect of Temperature on the Intrinsic Flexibility of DNA and Its Interaction with Architectural Proteins. *Biochemistry* **2014**, *53*, 6430–6438.

(61) Mouhat, F.; Peria, M.; Morresi, T.; Vuilleumier, R.; Saitta, A. M.; Casula, M. Thermal dependence of the hydrated proton and optimal proton transfer in the protonated water hexamer. *Nat. Commun*. **2023**, *14*, 6930.

(62) Scheiner, S.; Kern, C. W. Molecular orbital investigation of multiply hydrogen bonded systems. Formic acid dimer and DNA base pairs. *J. Am. Chem. Soc.* **1979**, *101*, 4081–4085.

(63) Winokan, M.; Slocombe, L.; Al-Khalili, J.; Sacchi, M. Multiscale simulations reveal the role of PcrA helicase in protecting against spontaneous point mutations in DNA. *Sci. Rep.* **2023**, *13*, 21749.

(64) Fillaux, F.; Cousson, A.; Gutmann, M. J. Proton transfer across hydrogen bonds: From reaction path to Schrödinger's cat. *Pure Appl. Chem.*, **2007**, *79*, 1023–1039.

(65) Fillaux, F.; Cousson, A.; Gutmann, M. J. Macroscopic quantum entanglement and 'super-rigidity' of protons in the KHCO3 crystal from 30 to 300 K. *J. Phys.: Cond. Matter* **2006**, *18*, 3229.

(66) Gorb, L.; Podolyan, Y.; Leszczynski, J.; Siebrand, W.; Fernández-Ramos, A.; Smedarchina, Z. A quantum-dynamics study of the prototropic tautomerism of guanine and its contribution to spontaneous point mutations in Escherichia coli. *Biopolymers* **2001**, *61*, 77–83.

(67) Waluk, J. Nuclear Quantum Effects in Proton or Hydrogen Transfer. *J. Phys. Chem. Lett.* **2024**, *15*, 598–607.

(68) Douhal, A.; Kim, S. K.; Zewail, A. H. Femtosecond molecular dynamics of tautomerization in model base pairs. *Nature* **1995**, *378*, 261-263.

(69) Fillaux, F.; Cousson, A. A neutron diffraction study of the crystal of benzoic acid from 6 to 293 K and a macroscopic-scale quantum theory of the lattice of hydrogen-bonded dimers. *Chem. Phys.* **2016**, *479*, 26–35.

(70) Nakanishi, T.; Hori, Y.; Shigeta, Y.; Sato, H.; Wu, S.-Q.; Kiyanagi, R.; Munakata, K.; Ohharad, T.; Sato, O. Observation of proton-transfer-coupled spin transition by single-crystal neutron-diffraction measurement. *Phys. Chem. Chem. Phys.* **2023**, *25*, 12394-12400.

(71) Lancaster, T.; Blundell, S. J.; Baker, P. J.; Brooks, M. L.; Hayes, W.; Pratt, F. L.; Manson, J. L.; Conner, M. M.; Schlueter, J. A. Muon-Fluorine Entangled States in Molecular Magnets. *Phys. Rev. Lett.* **2007**, *99*, 267601.

(72) Lin, L.; Morrone, J. A.; Car, R. Correlated Tunneling in Hydrogen Bonds. *J. Stat. Phys.* **2011**, *145*, 365–384.

(73) Ceriotti, M.; Cuny, J.; Parrinello, M.; Manolopoulos, D. E. Nuclear quantum effects and hydrogen bond fluctuations in water. *Proc. Natl. Acad. Sci. U.S.A.* **2013**, *110*, 15591–15596.

(74) Jiang, J.; Gao, Y.; Li, L.; Liu, Y.; Zhu, W.; Zhu, C.; Francisco, J. S.; Zeng, X. C. Rich proton dynamics and phase behaviours of nanoconfined ices. *Nat. Phys.* **2024**, *20*, 456–464.

(75) Matusalem, F.; Rego, J. S.; de Koning, M. Plastic deformation of superionic water ices. *Proc. Natl. Acad. Sci. U.S.A.* **2022**, *119*, e2203397119.

(76) Soper, A. K.; Benmore, C. J. Quantum Differences between Heavy and Light Water. *Phys. Rev. Lett.* **2008**, *101*, 065502.

(77) Morrone, J. A.; Lin, L.; Car, R. Tunneling and delocalization effects in hydrogen bonded systems: A study in position and momentum space. *J. Chem. Phys*. **2009**, *130*, 204511.

(78) De Brito Mota, F.; Rivelino, R. Properties of cyclo-β-tetrapeptide assemblies investigated by means of DFT calculations. *J. Mol. Struct.: Theochem* **2008**, *776*, 53–59.

(79) Clary, D. C.; Benoit, D. M.; van Mourik, T. H-Densities: A New Concept for Hydrated Molecules. *Acc. Chem. Res*. **2000**, *33*, 441–447.

(80) Xu, J.; Zhou, R.; Tao, Z.; Malbon, C.; Blum, V.; Hammes-Schiffer, S.; Kanai, Y. Nuclear–electronic orbital approach to quantization of protons in periodic electronic structure calculations. *J. Chem. Phys*. **2022**, *156*, 224111.

(81) Sirjoosingh, A.; Pak, M. V.; Brorsen, K. R.; Hammes-Schiffer, S. Quantum treatment or protons with the reduced explicitly correlated Hartree Fock approach. *J. Chem. Phys*. **2015**, *142*, 214107.

(82) Leggett, A. J. Testing the limits of quantum mechanics: motivation, state of play, prospects. *J. Phys.: Condens. Matter* **2002**, *14*, R415.

(83) Casanova, D. Theoretical Modeling of Singlet Fission. *Chem. Rev*. **2018**, *118*, 7164-7207.

(84) McWeeny, R. Methods of molecular quantum mechanics. Academic Press: London, 1989.

(85) Scheiner, S. Theoretical Studies of Proton Transfers. *Acc. Chem. Res*. **1985**, *18*, 174–180.

(86) Wang, R.; Zhang, C.; Zhang, B.; Liu, Y.; Wang, X.; Xiao, M. Magnetic dipolar interaction between correlated triplets created by singlet fission in tetracene crystals. *Nat. Commun.* **2015**, *6*, 8602.

(87) Nagashima, K.; Velan, S. S. Understanding the singlet and triplet states in magnetic resonance. *Concepts Magn. Reson*. **2013**, *42* 165-181.

(88) Zhou, Y.; Han, S. T.; Chen, X.; Wang, F.; Tang, Y.-B.; Roy, V. A. L. An upconverted photonic nonvolatile memory. *Nat. Commun.* **2014**, *5*, 4720.

(89) Smyser, K. E.; Eaves, J. D. Singlet fission for quantum information and quantum computing: the parallel JDE model. *Sci. Rep*. **2020**, *10*, 18480.

(90) Yong, C.; Musser, A.; Bayliss, S.; *et al*. The entangled triplet pair state in acene and heteroacene materials. *Nat. Commun*. **2017**, *8*, 15953.

(91) Stovbun, S. V.; Zlenko, D. V.; Bukhvostov, A. A.; Vedenkin, A. S.; Skoblin, A. A.; Kuznetsov, D. A.; Buchachenko, A. L. Magnetic field and nuclear spin influence on the DNA synthesis rate. *Sci. Rep*. **2023**, *13*, 465.

(92) Goldanskii, V.; Trakhtenberg, L.; Fleurov, V. Tunneling Phenomena in Chemical Physics. Taylor & Francis: London, 1989.

(93) Greve, C.; Elsaesser, T. Ultrafast Two-Dimensional Infrared Spectroscopy of Guanine–Cytosine Base Pairs in DNA Oligomers. *J. Phys. Chem. B* **2013** *117*, 14009–14017.

(94) Del Amo, J. M. L.; Langer, U.; Torres, V.; Buntkowsky, G.; Vieth, H.-M.; Pérez-Torralba, M.; Sanz, D.; Claramunt, R. M.; Elguero, J.; Limbach, H.-H. NMR Studies of Ultrafast Intramolecular Proton Tautomerism in Crystalline and Amorphous N,N′-Diphenyl-6-aminofulvene-1-aldimine: Solid-State, Kinetic Isotope, and Tunneling Effects. *J. Am. Chem. Soc*. **2008**, *130*, 8620-8632.

(95) Burdett, J. J; Bardeen, C. J. Quantum Beats in Crystalline Tetracene Delayed Fluorescence Due to Triplet Pair Coherences Produced by Direct Singlet Fission. *J. Am. Chem. Soc*. **2012**, *134*, 8597−8607.

(96) Piland, G. B.; Burdett, J. J.; Dillon, R. J.; Bardeen, C. J. Singlet Fission: From Coherences to Kinetics. *J. Phys. Chem. Lett*. **2014**, *5*, 2312−2319.

(97) Kandrashkin, Y. E.; Wang, Z.; Sukhanov, A. A.; Hou, Y.; Zhang, X.; Liu, Y.; Voronkova, V. K.; Zhao, J. Balance between Triplet States in Photoexcited Orthogonal BODIPY Dimers. *J. Phys. Chem. Lett*. **2019**, *10*, 4157−4163.

(98) Difley, S.; Beljonne, D.; Van Voorhis, T. On the Singlet−Triplet Splitting of Geminate Electron−Hole Pairs in Organic Semiconductors. *J. Am. Chem. Soc*. **2008**, *130*, 3420-3427.

(99) Climente, J. I. Origin of two-hole triplet splitting in circular quantum dots. *Solid State Commun*. **2012**, *152*, 825-829.

(100) Benk, H.; Sixl, H. Theory of 2 coupled triplet-states application to bicarbene structures. *Mol. Phys*. **1981**, *42*, 779–801.

(101) Picazo-Frutos, R.; Sheberstov, K. F.; Blanchard, J. W.; Van Dyke, E.; Moritz Reh, M.; Tobias Sjoelander, T.; Pines, A.; Budker D.; Barskiy, D. A. Zero-field j-spectroscopy of quadrupolar nuclei. *Nat. Commun*. **2024**, *15*, 4487.

(102) Jones, J. A.; Mosca, M. Implementation of a quantum algorithm on a nuclear magnetic resonance quantum computer. *J. Chem. Phys*. **1998**, *109*, 1648 – 1653.

(103) Martínez, F. A.; Aucar, G. A.; Intermolecular magnetic interactions in stacked DNA base pairs. *Phys. Chem. Chem. Phys*. **2017**, *19*, 27817–27827.

(104) Belousov, Y.; Chernousov, I.; Man'ko, V. Pseudo-Qutrit Formed by Two Interacting Identical Spins ($s = 1/2$) in a Variable External Magnetic Field. *Entropy* **2023**, *25*, 716.

(105) Nguyen, V. H. Quantum dynamics of two-spin-qubit systems. *J. Phys. Condens. Matter* **2009**, *21*, 273201.

(106) Loss, D.; DiVincenzo, D. P. Quantum computation with quantum dots. *Phys. Rev. A* **1998**, *57*, 120–126.

(107) Goss, N.; Ferracin, S.; Hashim, A.; Carignan-Dugas, A.; Kreikebaum, J. M.; Naik, R. K.; Santiago, D. I.; Siddiqi, I. Extending the computational reach of a superconducting qutrit processor. *npj Quantum Inf*. **2024**, *10*, 101.

(108) Hu, X.; Das Sarma, S. Hilbert-space structure of a solid-state quantum computer: Two-electron states of a double-quantum-dot artificial molecule. *Phys. Rev. A* **2000**, *61*, 062301.

(109) Burkard, G.; Loss, D.; DiVincenzo, D. P. Coupled quantum dots as quantum gates. *Phys. Rev. B* **1999**, *59*, 2070.

(110) Wong, T. G. Introduction to Classical and Quantum Computing. Rooted Grove: Nebraska 2022. This excellent textbook is available as a free ebook (https://www.thomaswong.net/introduction-to-classical-and-quantum-computing-1e4p.pdf).

(111) Seeley, J. T.; Richard, M. J.; Love, P. J. The Bravyi-Kitaev transformation for quantum computation of electronic structure. *J. Chem. Phys*. **2012**, *137*, 224109.

(112) Bravyi, S; Kitaev, A. Fermionic Quantum Computation. *Ann. Phys*. **2002**, *298*, 210.

(113) O'Malley, P. J. J.; Babbush, R.; Kivlichan, I. D.; *et al*. Scalable Quantum Simulation of Molecular Energies. *Phys. Rev. X* **2016**, *6*, 031007.

(114) Riera Aroche, R.; Ortiz García, Y. M.; Martínez Arellano, M. A.; Riera Leal, A. DNA as a perfect quantum computer based on the quantum physics principles. *Sci. Rep*. **2024**, *14*, 11636.

(115) Ladd, T. D.; Jelezko, F.; Laflamme, R.; Nakamura, Y.; Monroe, C.; O'Brien, J. L. Quantum computers. *Nature* **2010**, *464*, 45–53.

(116) Gao, A.; Remsing, R. C. Self-consistent determination of long-range electrostatics in neural network potentials. *Nat. Commun*. **2022**, *13*, 1572.

(117) Shee, Y.; Yeh, T. L.; Hsiao, J. Y.; Yang, A.; Yen-Chu Lin, Y.-C.; Hsieh, M.-H. Quantum simulation of preferred tautomeric state prediction. *npj Quantum Inf*. **2023**, *9*, 102.

(118) Peruzzo, A.; McClean, J.; Shadbolt, P.; Yung, M.-H.; Zhou, X.-Q.; Love, P. J.; Aspuru-Guzik, A.; Jeremy L. O'Brien, J. L. A variational eigenvalue solver on a photonic quantum processor. *Nat. Commun*. **2014**, *5*, 4213.

(119) Abrams, D. S.; Lloyd, S. Simulation of Many-Body Fermi Systems on a Universal Quantum Computer. *Phys. Rev. Lett*. **1997**, *79*, 2586–2589.

(120) Berry, D. W.; Kieferová, M.; Scherer, A.; Sanders, Y. R.; Low, G. H.; Wiebe, N.; Gidney, C.; Babbush, R. Improved techniques for preparing eigenstates of fermionic Hamiltonians. *npj Quantum Inf*. **2018**, *4*, 22.

(121) Zhao, L.; Goings, J.; Shin, K.; Kyoung, W.; Fuks, J. I.; Rhee, J.-K. K.; Rhee, Y. M.; Wright, K.; Nguyen, J.; Kim, J.; Johri, S. Orbital-optimized pair-correlated electron simulations on trapped-ion quantum computers. *npj Quantum Inf*. **2023**, *9*, 60.

(122) Moll, N.; Barkoutsos, P.; Bishop, L. S. *et al*. Quantum optimization using variational algorithms on near-term quantum devices. *Quantum Sci. Technol*. **2018**, *3* 030503.

(123) Dasgupta, K.; Paine, B. Loading Probability Distributions in a Quantum circuit. arXiv:2208.13372v1 [quant-ph].

(124) Vandersypen, L.; Steffen, M.; Breyta, G; Yannoni, C. S.; Sherwood, M. H.; Chuang, I. L. Experimental realization of Shor's quantum factoring algorithm using nuclear magnetic resonance. *Nature* **2001**, *414*, 883–887.

(125) Price, M. D.; Somaroo, S. S.; Tseng, C. H.; Gore, J. C.; Fahmy, A. F.; Havel, T. F.; Cory, D. G. Construction and Implementation of NMR Quantum Logic Gates for Two Spin Systems. *J. Magn. Reson.* **1999**, *140*, 371–378.

(126) Dong, D.; Wu, C.; Chen, C; Qi, B.; Petersen. I. R.; Nori, F. Learning robust pulses for generating universal quantum gates. *Sci. Rep*. **2016**, *6*, 36090.

(127) Kelly, J.; R. Barends, R.; B. Campbell, B. *et al*. Optimal Quantum Control Using Randomized Benchmarking. *Phys. Rev. Lett*. **2014**, *112*, 240504.

(128) Yan, B.; Moses, S. A.; Gadway, B.; Covey, J. P.; Hazzard, K. R. A.; Rey, A. M.; Jin, D. S.; Ye, J. Observation of dipolar spin-exchange interactions with lattice-confined polar molecules. *Nature* **2013**, 501, 521–526.

(129) Atzori, M.; Chiesa, A.; Morra, E.; Chiesa, M.; Sorace, L.; Carretta, S.; Sessoli. R. A two-qubit molecular architecture for electron-mediated nuclear quantum simulation. *Chem. Sci*. **2018**, *9*, 6183-6192.

(130) Bao, Y.; Yu, S. S.; Anderegg, L.; *et al*. Dipolar spin-exchange and entanglement between molecules in an optical tweezer array. *Science* **2023**, *382*, 1138-1143.

(131) Holland, C. M.; Lu, Y.; Cheuk, L. W. On-demand entanglement of molecules in a reconfigurable optical tweezer array. *Science* **2023**, *382*, 1143-1147.

(132) Stemp, H. G.; Asaad, S.; van Blankenstein, M. R.; *et al*. Tomography of entangling two-qubit logic operations in exchange-coupled donor electron spin qubits. *Nat. Commun.* **2024**, *15*, 8415.

(133) Choi, J.; Zhou, H.; Knowles, H. S.; Landig, R.; Choi, S.; Lukin, M. D. Robust Dynamic Hamiltonian Engineering of Many-Body Spin Systems. *Phys. Rev. X* **2020**, *10*, 031002.